\documentclass[pra,aps,final,twocolumn,showpacs,showkeys]{revtex4-2}
\usepackage{amsmath}
\usepackage{graphicx}
\usepackage{amsfonts}
\usepackage{amssymb}
\usepackage{subfigure}

\def\identity{\leavevmode\hbox{\small1\kern-3.2pt\normalsize1}}

\begin{document}

\title{Dephasing-induced leakage in multi-level superconducting quantum circuits}

\author{Frederick W. Strauch}
\email[Electronic address: ]{Frederick.W.Strauch@williams.edu}
\affiliation{Department of Physics, Williams College, Williamstown, MA 01267}

\date{\today}

\begin{abstract}
Superconducting quantum circuits, such as the transmon, have multiple quantum states beyond the computational subspace.  These states can be populated during quantum logic operations; residual population in such states is known as leakage.  While control methods can eliminate this error in ideal systems, in the presence of dephasing transient population leads to leakage.  This dephasing-induced leakage effect is analyzed, both analytically and numerically, for common single and two-qubit operations used in transmon-based approaches to quantum information processing.    \end{abstract} 
\maketitle

\section{Introduction}
Superconducting qubits have progressed from fundamental proofs-in-principle experiments in coherence \cite{martinis2002rabi,vion2002manipulating,chiorescu2003coherent} and entanglement \cite{pashkin2003quantum,berkley2003entangled,yamamoto2003demonstration} twenty years ago to a leading platform for quantum information processing, most recently through the recent demonstration of magic state preparation by IBM Quantum \cite{gupta2024encoding} and surface code quantum error correction by Google Quantum AI \cite{google2024qec}.  Nevertheless, obstacles to large-scale quantum computation may remain.  Previous work by Google showed \cite{google2023suppressing} that leakage, the population of quantum states outside of the standard computational basis, is one such obstacle.  As such, understanding the mechanisms behind leakage and how to mitigate this error is an important task.  Indeed, the continued progress of superconducting qubits has relied on control methods used to minimize leakage in single \cite{chen2016measuring} and two-qubit \cite{barends2014superconducting,kelly2014optimal} operations, and the recent demonstration of quantum error correction used leakage reduction methods \cite{miao2023overcoming}.  

In this paper I consider a fundamental mechanism that limits quantum control techniques for leakage elimination.  This mechanism is the dephasing of superpositions between qubit and non-qubit states during gate operations.  This occurs during microwave-driven single-qubit gates, in which there is transient population of the second excited state, as shown in Fig.~1(a).  This also occurs during two-qubit gates that rely on the interaction between $|11\rangle$ and $|20\rangle$ states, as shown in Fig.~1(b).  The dephasing-induced leakage is calculated, both analytically and numerically, for these two types of operations.  For white noise, the leakage probability scales as $T/T_{\varphi}$, where $T$ is the gate time, $T_{\varphi}$ is the single-qubit dephasing time, and ranges from $10^{-6}$ to $10^{-4}$ for the gates used in recent experiments \cite{google2023suppressing}, while low frequency noise leads to smaller amounts of leakage.  Better understanding of these control errors could guide future experiments.

\begin{figure}
\includegraphics[width=3 in]{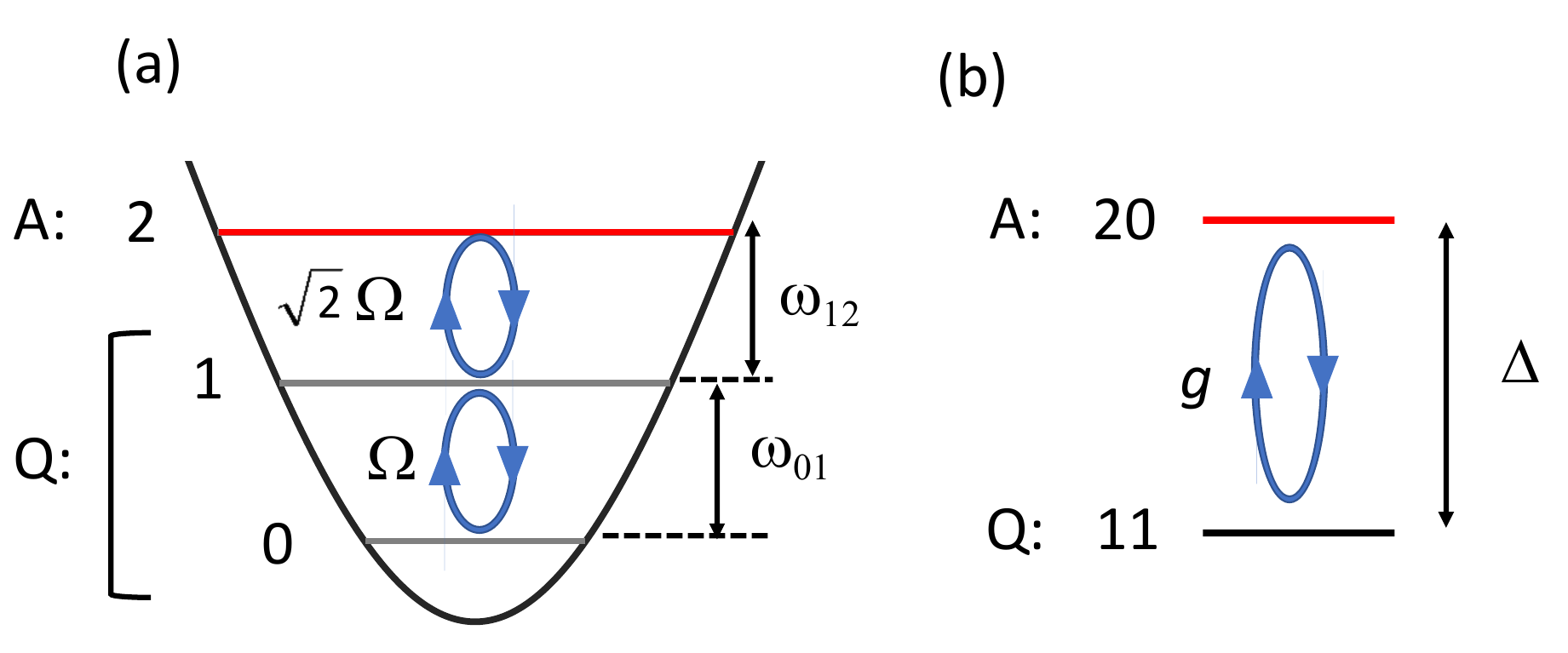} 
\label{fig1}
\caption{Coupling of qubit subspace $Q$ to an auxiliary subspace $A$ for qubit operations in a transmon circuit.  (a) Single-qubit operations with $Q = \{ |0\rangle, |1\rangle \}$ and $A = \{ |2\rangle \}$. Microwave driving at frequency $\omega_{01}$ causes Rabi osci6llations between $|0\rangle$ and $|1\rangle$ with frequency $\Omega$, as indicated by the cycle between energy levels.  This drive also couples to state $|2\rangle$ with amplitude $\sqrt{2} \Omega$ with a detuning of $\omega_{01}-\omega_{12} = \eta$.  This causes transient population of $|2\rangle$ during gate operations.  (b). Two-qubit controlled-phase gate with $Q = \{|11\rangle \}$ and $A = \{ |20\rangle \}$.  For fixed coupling $g$, rapid or adiabatic tuning of $\Delta = \omega_{12,1} - \omega_{01,2}$ can generates a net phase on state $|1\rangle_1 |1\rangle_2$ (for transmon $1$ and transmon $2$).  For tunable coupling, a $2 \pi$ rotation yields $|11\rangle \to -i |20\rangle \to -|11\rangle$. }
\end{figure}

This paper is organized as follows.  In Sec. II I provide a general perturbative framework to calculate leakage due to dephasing caused by classical noise processes with arbitrary spectral densities.  Specific calculations for white noise and $1/f$ noise process are presented for the two-qubit controlled-phase gate in Sec. III, and for a single-qubit NOT gate in Sec. IV.  The white noise calculations are in excellent agreement with master equation simulations for the density matrix.  These simulations incorporate control methods commonly used to achieve low leakage in the absence of dephasing.  The noise and simulation details are described in the Appendices.  The paper concludes in Sec. V with observations on possible research directions.

\section{General Leakage Framework}

The general problem is the following.  Consider a quantum system with a finite-dimensional Hilbert space that is separated into two subspaces.  The first subspace is a computational, or qubit, subspace $Q$ of dimension $d_Q$, spanned by the states $|q_j\rangle$, $j=1 \to d_Q$.  The second is an auxiliary subspace $A$ of dimension $d_A$, spanned by states $|a_k\rangle$, $k=1 \to d_A$.  Let $P_Q$ and $P_A$ be the corresponding projectors onto those subspaces, such that $P_Q + P_A = I$.  The Hamiltonian for this system is assumed to be of the form
\begin{equation}
\mathcal{H} = \mathcal{H}_{0} + \mathcal{H}_{\text{noise}},
\end{equation}
where $\mathcal{H}_{0}$ is an ideal, controllable Hamiltonian, while $\mathcal{H}_{\text{noise}}$ is a noisy, uncontrolled Hamiltonian.  The desired operation is $U_{0}(T)$, where the time-evolution operator is
\begin{equation}
U_{0}(t) = \mathcal{T} \exp \left[- i \int_0^t \mathcal{H}_{0}(t') dt' \right]
\end{equation}
and $\mathcal{T}$ is the time-ordering symbol (and $\hbar$ is included in the definition of $\mathcal{H}$).  For the operations of interest, $U_0(T)$ commutes with the subspace projectors $P_Q$ and $P_A$, but $U_0(t)$ does not.  That is, there will be transient population in the auxiliary states.  It remains to calculate how $\mathcal{H}_{\text{noise}}$ causes leakage into the auxiliary subspace.  

To isolate the effect of noise, I use an interaction picture and let the full time-evolution operator be $U(t) = U_{0}(t) U_{1}(t)$, so that
\begin{equation}
U_{1}(t) = \mathcal{T} \exp \left[- i \int_0^t \mathcal{H}_{1}(t') dt' \right],
\end{equation}
where $\mathcal{H}_1$ is the interaction-picture Hamiltonian
\begin{equation}
\mathcal{H}_1 = U_0(t)^{\dagger} \mathcal{H}_{\text{noise}} U_0(t).
\end{equation}
The resulting leakage probability, when averaged over all initial states of the qubit subspace, is given by
\begin{align}
\mathcal{P}_{\text{leakage}} &= \frac{1}{d_Q} \text{Tr} \left[ P_A U(T) P_Q U(T)^{\dagger} \right] \nonumber \\
&= \frac{1}{d_Q} \text{Tr} \left[ P_A U_1(T) P_Q U_1(T)^{\dagger} \right] \nonumber \\
& = \frac{1}{d_Q} \sum_{j=1}^{d_Q} \sum_{k=1}^{d_A}| \langle a_k | U_1(T) | q_j \rangle |^2.
\end{align}
and I have used the assumption that $U_0(T)$ commutes with the subspace projectors and the cyclic property of the trace to obtain the second line.

To proceed, I assume that the noise Hamiltonian represents a dephasing process, and can be written as
\begin{equation}
\mathcal{H}_{\text{noise}}(t) = \varepsilon(t) V_0.
\end{equation}
Here $V_0$ is a time-independent diagonal matrix in the qubit and auxiliary state basis, and the stochastic process $\varepsilon(t)$ has the correlation function
\begin{equation}
C(t-t')  = \langle \varepsilon(t) \varepsilon(t') \rangle_{\text{noise}} = \frac{1}{2\pi} \int_{-\infty}^{\infty} S(\omega) e^{i \omega (t-t')} d \omega,
\end{equation}
with spectral density $S(\omega)$.  Two types of spectral density functions will be considered in this paper.  The first is white noise $S(\omega) = S_0$, which gives rise to single-qubit dephasing of the exponential form $\exp[-t/T_{\varphi}^{(1)}]$.  The second is low frequency ($1/f$) noise, with $S(\omega) = S_1/|\omega|$ (with a low frequency cutoff).  As discussed in Appendix A, this gives rise to single-qubit dephasing of the Gaussian form $\exp[-(t/T_{\varphi}^{(2)})^2]$.
I now use time-dependent perturbation theory for $U_1$ to find
\begin{equation}
U_1(T) \approx I - i \int_{0}^T \varepsilon(t) V_1(t) dt,
\end{equation}
where $V_1(t) = U_0(t)^{\dagger} V_0 U_0(t)$.  Using this approximation, the noise-averaged leakage probability can be written as
\begin{align}
\mathcal{P}_\text{leakage}  &= \frac{1}{d_Q} \sum_{j=1}^{d_Q} \sum_{k=1}^{d_A} \int_0^T dt \int_0^T dt' C(t-t') \mathcal{A}_{j \to k}(t) \mathcal{A}_{j \to k}^* (t') \nonumber \\
&= \frac{1}{d_Q}  \sum_{j=1}^{d_Q} \sum_{k=1}^{d_A} \int_{-\infty}^{\infty} \frac{1}{2\pi} S(\omega) |\tilde{\mathcal{A}}_{j \to k}(\omega)|^2 d \omega,
\label{correlatedL}
\end{align}
where 
\begin{equation}
\mathcal{A}_{j \to k}(t)  = \langle a_k| V_1(t) | q_j \rangle
\end{equation} 
and
\begin{equation}
\tilde{\mathcal{A}}_{j \to k} (\omega) = \int_0^T e^{ i \omega t'} \mathcal{A}_{j \to k}(t') dt'.
\end{equation}

 For the gates considered in this paper, the functions $|\tilde{\mathcal{A}}_{j \to k}(\omega)|^2$ are peaked about a characteristic frequency $\omega_{j \to k}$.  This allows a further simplification by replacing $S(\omega)$ by $S(\omega_{j \to k})$ in Eq.~(\ref{correlatedL}), so that
\begin{align}
\mathcal{P}_{\text{leakage}} &\approx \frac{1}{d_Q}  \sum_{j=1}^{d_Q} \sum_{k=1}^{d_A} S(\omega_{j \to k}) \int_{-\infty}^{\infty} \frac{1}{2\pi}|\tilde{\mathcal{A}}_{j \to k}(\omega)|^2 d \omega \nonumber \\
& = \frac{1}{d_Q} \sum_{j=1}^{d_Q} \sum_{k=1}^{d_A} S(\omega_{j \to k}) \int_{0}^T | \mathcal{A}_{j \to k}(t) |^2 dt.
\label{generalL}
\end{align}
In this form, the total leakage probability is given by the time average of the instantaneous transition probabilities $|\mathcal{A}_{j \to k}(t)|^2$.  While the amplitudes $\mathcal{A}_{j \to k}(t)$ can be optimized to vanish at times $t=0$ and $t=T$, for the cases under consideration, they do not vanish at intermediate times $0 < t < T$.  Thus, transient population of auxiliary states, when subject to noise, leads to a net leakage of probability.

This is a very general result.   If $U_0(t)$ has non-zero matrix elements between the qubit and auxiliary subspaces during the time interval $0 < t < T$, dephasing noise causes leakage.  This is true even if $\mathcal{H}_{\text{noise}}$ is diagonal in these subspaces, as the interaction picture Hamiltonian $\mathcal{H}_1(t) = U_0(t)^{\dagger} \mathcal{H}_{\text{noise}} U_0(t)$ will typically not be diagonal.  Thus, the total evolution will lead to some probability of leakage.  How much leakage will be the subject of the following sections. 

\section{Two-Qubit Controlled-Phase Gate}

Our first operation is the two-qubit controlled-phase gate between transmons.  The original proposal \cite{strauch2003quantum}, analyzed using phase qubits, involves a stimulated $2 \pi$ rotation of the two-qubit state $|q\rangle = |11\rangle$ with the auxiliary quantum state $|a\rangle = |20\rangle$, of the form $|11\rangle \to -i |20\rangle \to - |11\rangle$.  As the other states are unchanged, one can analyze this using a two-state Hamiltonian of the form
\begin{equation}
\mathcal{H}_0 = \left( \begin{array}{cc} 0 & g \\ g & \Delta \end{array} \right),
\label{phase_hamiltonian}
\end{equation}
where $g$ is the coupling between states and the detuning $\Delta = \omega_{01,1} - \omega_{12,2}$, where $\omega_{01,1}$ is the transition frequency of transmon $1$, $\omega_{01,2}$ is the transition frequency of transmon $2$, $\omega_{12,2} = \omega_{01,2} - \eta_2$, and $\eta_2$ is the anharmonicity of transmon $2$.  The noise Hamiltonian is given by $\mathcal{H}_{\text{noise}} = \varepsilon(t) V_0$ with
\begin{equation}
V_0 = \begin{pmatrix} 0 & 0 \\ 0 & 1 \end{pmatrix}.
\end{equation}
The noise process $\varepsilon(t)$ is taken to be the sum of two independent noise processes (one for each qubit).

The controlled-phase gate can be implemented in a number of ways.  For fixed coupling, one can perform a rapid tuning \cite{strauch2003quantum} of the qubit frequencies so that $\Delta=0$ for an interaction time $T = \pi/g$.  This frequency tuning can also be performed quasi-adiabatically, as demonstrated in \cite{dicarlo2009demonstration}.  Variations of this method include the ``fast adiabatic'' gate used in \cite{barends2014superconducting,kelly2014optimal}.  Another notable variant uses an echo-based approach to minimize errors associated with frequency tuning \cite{rol2019fast,negirneac2021high}; numerical simulations of their ``net zero'' gate indicated that leakage was associated with dephasing.  

For tunable coupling $g(t)$, there are two typically two regimes.  For a slowly varying coupler \cite{chen2014qubit}, one can operate in a small detuning regime ($\Delta < g$) so that $\int_0^T g(t) dt = \pi$.  For a parametric coupling $g = g_0 \cos(\omega t)$, one can operate in the large detuning regime ($\Delta > g_0$) with drive frequency $\omega = \Delta$ \cite{ganzhorn2020benchmarking, jin2023versatile}.  One can also parametrically activate the gate by modulating the frequency of one of the qubits \cite{caldwell2018parametrically,reagor2018demonstration}.  

For simplicity, I will consider only the fixed coupling operations using rapid or adiabatic tuning.  Similar results can be expected for other gate implementations that use this type of interaction between $|11\rangle$ and $|20\rangle$.

\subsection{Rapid Tuning}

In the rapid tuning operation, one sets $\Delta = 0$ in $\mathcal{H}_0$, so that
\begin{equation}
U_0(t) = \begin{pmatrix} \cos(g t) & -i \sin(g t) \\[0.1cm]-i \sin(g t) & \cos(g t) \end{pmatrix}.
\end{equation}
The noise operator $V_0$ transforms to 
\begin{equation}
V_1(t) = U_0(t)^{\dagger} V_0 U_0(t) = \begin{pmatrix} \sin^2 (g t) & i \sin(2 g t)/2 \\[0.1cm] -i \sin(2 g t)/2 & \cos^2 (g t) \end{pmatrix},
\end{equation}
and thus 
\begin{equation}
\mathcal{A}(t) = \langle a | V_1(t) |q\rangle =  -i \frac{1}{2} \sin(2 g t).
\end{equation}
This amplitude leads to 
\begin{equation}
\tilde{\mathcal{A}}(\omega) = \frac{2 g}{\omega^2 - 4 g^2} e^{i \omega T/2} \sin(\omega T/2)
\end{equation}
with $T=\pi/g$, which is peaked at $\omega_0 \approx 1.7 g$.  Using Eq.~(\ref{generalL}) with $d_Q = d_A = 1$, the dephasing-induced leakage can be approximated by
\begin{equation}
\mathcal{P}_{\text{leakage}} \approx \frac{1}{4} S(\omega_0) \int_{0}^T \sin^2 (2 g t) dt = \frac{1}{8} S(\omega_0) T.
\end{equation}

For white-noise, this approximation is exact.  The appropriate spectral density for this two-qubit problem is $S_0 = 4/T_{\varphi}^{(1)}$, where $T_{\varphi}^{(1)}$ is the single-qubit dephasing time.  Using this value of $S_0$ results in
\begin{equation}
\mathcal{P}_{\text{leakage}}^{(1)} \approx \frac{1}{2} \left( \frac{T}{T_{\varphi}^{(1)}} \right).
\label{rapidL}
\end{equation}
This result agrees with solutions of the master equation for the density matrix (see Appendix B).  For physical parameters $g/(2\pi) = 50 \ \mbox{MHz}$ (so that $T = \pi / g = 10 \ \mbox{ns}$) and $T_{\varphi}^{(1)} = 100 \ \mu \text{s}$, the leakage probability is $\mathcal{P}_{\text{leakage}}^{(1)} \approx 5 \times 10^{-5}$.  For comparison, the equivalent leakage probability quoted in \cite{google2023suppressing} is $8 \times 10^{-4}$ (as discussed in Sec. IV of their Supplementary Information).  However, the gate performed there uses a tunable coupler and has a longer gate duration of $30 \ \mbox{ns}$.  Thus, the leakage calculated here can be considered as an approximate lower bound in the presence of qubit dephasing.

For $1/f$ noise, one can parametrize the spectral density via $S_1 = (2 / 2.6) [T_{\varphi}^{(2)}]^{-2}$ (see Appendix A), and either evaluate the frequency integral in Eq.~(\ref{correlatedL}) numerically or use the approximation of Eq.~(\ref{generalL}).  Both result in the expression
\begin{equation}
\mathcal{P}_{\text{leakage}}^{(2)} \approx c  \left(\frac{T}{T_{\varphi}^{(2)}} \right)^2.
\label{rapidLlowf}
\end{equation}
The ``exact'' frequency integral yields $c = 0.021$, while the approximation with $\omega_0 \approx 1.675 g$ produces the smaller value $c = 0.018$.  For a gate time $T = 10 \ \mbox{ns}$ and a Gaussian decay time $T_{\varphi}^{(2)} = 1 \ \mu \text{s}$, $\mathcal{P}_{\text{leakage}}^{(2)} \approx 2 \times 10^{-6}$.  

\begin{figure*}
\includegraphics[width=6.5 in]{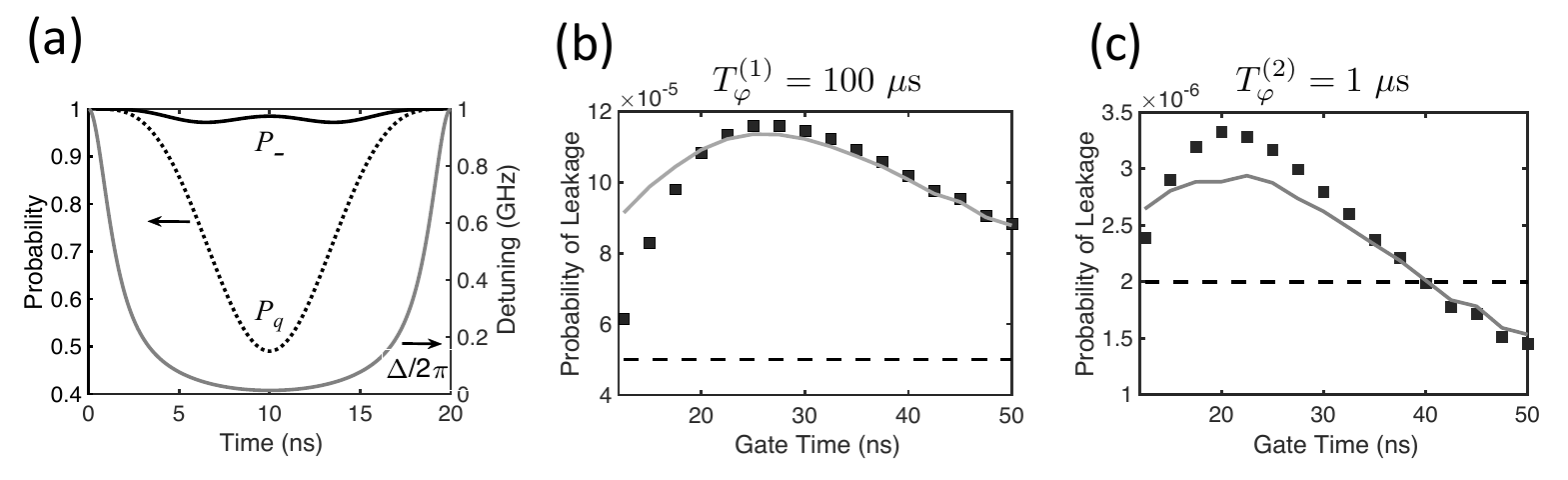}
\label{fig2}
\caption{(a) Optimized adiabatic tuning of the two-qubit controlled-phase gate.  The optimized trajectory for $\Delta(t)/2\pi$ is shown (lower solid curve, axis on right) given an initial detuning $\Delta(0)/2\pi = 1 \ \mbox{GHz}$, fixed coupling $g/2\pi = 50 \ \mbox{MHz}$, and gate time $T = 20 \ \mbox{ns}$.  Also shown are the simulated state probabilities (axis on left) $P_q(t) = |\langle q | \psi(t) \rangle|^2$ (dashed) and $P_{-}(t) = |\langle \psi_-(t) | \psi(t) \rangle|^2$ (solid), given the initial condition $|\psi(t=0)\rangle = |\psi_-(0)\rangle$ (see text). (b) Dephasing-induced leakage for white noise.  The probability of leakage $\mathcal{P}_{\text{leakage}}^{(1)}$ as a function of gate time $T$ calculated using Eq.~(\ref{adiabaticL}) (solid curve) or master equation simulations (squares), each using an optimized trajectory $\Delta(t)$ for each value of $T$ and for $T_{\varphi}^{(1)} = 100 \ \mu \mbox{s}$.  Also shown (dashed line) is the leakage found for rapid tuning, calculated using Eq.~(\ref{rapidL}).  (c) Dephasing-induced leakage for low-frequency noise.  The probability of leakage $\mathcal{P}_{\text{leakage}}^{(2)}$ as a function of gate time $T$ calculated using Eq.~(\ref{adiabaticL}) (solid curve) or by a numerical integration of Eq.~(\ref{correlatedL}) (squares), each using an optimized trajectory $\Delta(t)$ for each value of $T$ and for $T_{\varphi}^{(2)} = 1 \ \mu \mbox{s}$.  Also shown (dashed line) is the leakage found for rapid tuning, calculated using Eq.~(\ref{rapidLlowf}).}
\end{figure*}

\subsection{Adiabatic Tuning}
Since errors in timing can cause leakage in the rapid-tuning operation, an adiabatic implementation of this gate has been considered \cite{barends2014superconducting}.  For fixed coupling $g$, this requires the use of an appropriate frequency trajectory for $\Delta(t)$ so that the quantum state evolves according to the instantaneous eigenstate of $\mathcal{H}_0$ during the gate operation.  These eigenstates are
\begin{equation}
|\psi_{-}\rangle = \begin{pmatrix} \ \ \ \cos (\theta/2) \\ -\sin (\theta/2) \end{pmatrix} \ \mbox{and} \ \ |\psi_{+}\rangle = \begin{pmatrix} \sin (\theta/2) \\ \cos(\theta/2) \end{pmatrix}
\end{equation}
where 
\begin{equation}
\theta = \arctan \left( \frac{2 g}{\Delta} \right),
\end{equation}
with corresponding eigenvalues
\begin{equation}
E_{\pm} = \frac{1}{2} \Delta \pm \frac{1}{2} \sqrt{\Delta^2 + 4 g^2}.
\end{equation}
For an appropriately chosen trajectory $\Delta(t)$, the time-evolution operator can be approximated by
\begin{equation}
U_0(t) \approx e^{-i \phi_-(t)} |\psi_-(t)\rangle \langle \psi_-(0)| + e^{-i \phi_+(t)} |\psi_+(t)\rangle \langle \psi_+(0)|,
\label{adiabaticU}
\end{equation}
with
\begin{equation}
\phi_{\pm}(t) = \int_{0}^t E_{\pm}(t') dt'.
\end{equation}
Here $E_{\pm}(t)$ and $|\psi_{\pm}(t)\rangle$ are the instantaneous eigenvalues and eigenvectors of $\mathcal{H}_0$.   Note that the Berry phase \cite{berry1984quantal} is zero for this type of evolution. 

Martinis and Geller \cite{martinis2014fast} devised an approach to find trajectories for $\Delta(t)$ that lead to approximately adiabatic gates with relatively short gate times.  This approach has $\Delta(t)$ start and end with $\Delta(0) = \Delta(T) \gg g$ (where $\theta(0) \approx 0$), moving to $\Delta(T/2) \approx 0$ (where $\theta(T/2) \approx \pi/2$), and subject to the constraint $\phi_-(T) = \pi$.  Using Eq.~(\ref{adiabaticU}) for the approximate time-evolution operator, the noise operator transforms to
\begin{align}
V_1(t) & = U_0(t)^{\dagger} V_0 U_0(t) \nonumber \\
& = \begin{pmatrix} \sin^2 (\theta/2) & - e^{-i \phi} \sin(\theta) /2 \\[0.1cm]- e^{i \phi} \sin(\theta)/2 & \cos^2 (\theta/2) \end{pmatrix}
\end{align}
where I have suppressed the time-dependence of $\theta(t)=\arctan[2 g / \Delta(t)]$ and set
\begin{equation}
\phi(t) = \phi_+(t) -\phi_{-}(t).
\end{equation}
Thus, 
\begin{equation}
\mathcal{A}(t) = -\frac{1}{2} e^{i \phi(t)} \sin [\theta(t)].
\label{adiabaticA}
\end{equation}
The corresponding $\mathcal{A}(\omega)$ is peaked at a frequency $\omega_T$ that depends on the gate time $T$, so that
\begin{equation}
\mathcal{P}_{\text{leakage}} \approx \frac{1}{4} S(\omega_T) \int_{0}^T \sin^2 [\theta(t)] dt.
\label{adiabaticL}
\end{equation}

The leakage depends on the specific trajectory through $\theta(t)$, but can be numerically calculated for any given trajectory.  Inspired by \cite{martinis2014fast}, I use the parametrization $\Delta(t) = 2 g / \tan[ \theta(t)]$, with trajectories of the form 
\begin{equation}
\theta(t) = \theta_0 + \theta_1 [1 - \cos(2 \pi t/T)] + \theta_2 [1 - \cos(4 \pi t/T)],
\label{theta_trajectory}
\end{equation}
where $\tan \theta_0 = 2 g / \Delta(0)$ and $\theta_1$ and $\theta_2$ are found by optimizing numerical simulations of the time-dependent Schr{\"o}dinger equation so that $\langle \psi_-(0) | U_0(T) | \psi_-(0) \rangle = -1$.  A sample trajectory, with $\Delta(0)/2\pi = 1 \ \mbox{GHz}$ and $T=20 \ \mbox{ns}$, is shown in Fig.~2(a).  Also shown are the simulated state probabilities $P_q(t) = |\langle q | \psi(t) \rangle|^2$ and $P_{-}(t) = |\langle \psi_-(t) | \psi(t) \rangle|^2$, starting from the initial condition $|\psi(t=0)\rangle = |\psi_-(0)\rangle \approx |q\rangle$.  The dynamics is seen to be nearly adiabatic, as $P_-(t)$ remains close to unity, but the oscillation in $P_q(t)$ shows that there is a non-negligible transfer of population between $|q\rangle$ to $|a\rangle$ during the gate operation.

Repeating the optimization of $\theta(t)$ for each gate time, I use Eq.~(\ref{adiabaticL}) to calculate the leakage as a function of the gate time $T$.  This is shown in Fig.~2(b) for white noise (solid curve), along with the results of simulations (squares) of the corresponding master equation for the density matrix equation with single-qubit dephasing time $T_{\varphi}^{(1)} = 100 \ \mu \mbox{s}$ (as described in Appendix B).  The disagreement at smaller gate times arises from the adiabatic approximation used to obtain Eq.~(\ref{adiabaticA}).  A more accurate evaluation, using $\mathcal{A}(t)$ obtained from direct numerical simulation and integration of Eq.~(\ref{correlatedL}), agrees with the master equation results.  Also shown in Fig.~2(b) is the leakage from the rapid gate calculation of Eq.~(\ref{rapidL}) (dashed line), with $T = 10 \ \mbox{ns}$.   Somewhat surprisingly, the adiabatic tuning of the controlled-phase gate is seen to be more sensitive to leakage than the rapid operation, unless the gate time is made significantly longer ($T \sim 100 \ \mbox{ns}$).  

Note that, for fixed coupling $g$, the interaction is always on.  That is, in principle one can achieve a controlled-phase gate without any tuning.  For $\Delta \gg g$ we can approximate $\sin^2 \theta \approx 4 g^2/ \Delta^2$ and set $T = \pi/ (g^2/\Delta)$ in Eq.~(\ref{adiabaticL}), leading to 
\begin{equation}
\mathcal{P}_{\text{leakage}}^{(1)} \to \frac{4 \pi}{\Delta T_{\varphi}^{(1)}}.
\label{detuningL}
\end{equation}
This evaluates to $2 \times 10^{-5}$ for $\Delta/2\pi = 1 \ \mbox{GHz}$.  

The leakage for $1/f$ noise is shown in Fig.~2(c).   This is calculated from Eq.~(\ref{adiabaticL}) as the solid curve using the same optimized trajectories, and frequencies $\omega_T$ found from each $\mathcal{A}(t)$ using Eq.~(\ref{adiabaticA}), with $T_{\varphi}^{(2)} = 1 \ \mu \mbox{s}$.  Also shown is a more accurate calculation (squares) using direct numerical simulation to obtain $\mathcal{A}(t)$ and integration of Eq.~(\ref{correlatedL}).  Finally, and for comparison, the rapid gate result Eq.~(\ref{rapidLlowf}) is shown (dashed line).  These are similar to but consistently smaller than the corresponding white noise results.

\begin{figure*}
\includegraphics[width=6.5 in]{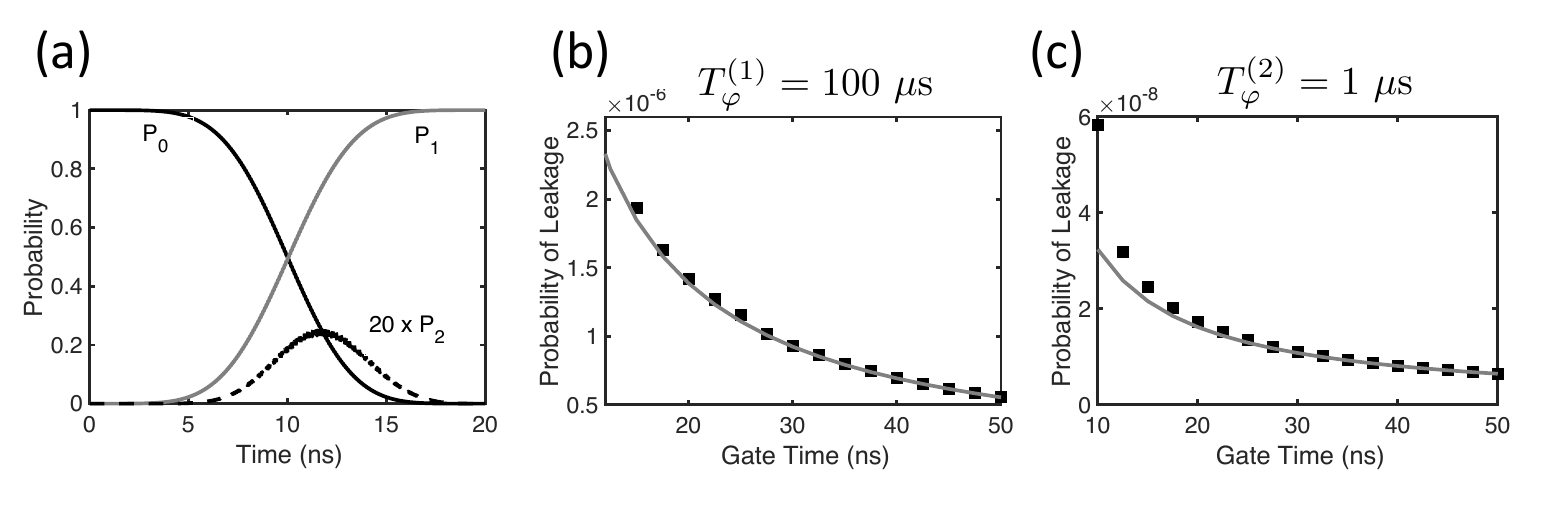}
\label{fig3}
\caption{(a) State probabilities $P_0(t) = |\langle 0 | \psi(t) \rangle|^2$, $P_1(t) = |\langle 1 |\psi(t) \rangle|^2$, and $P_2(t) = |\langle 2 |\psi(t)\rangle$ for a $T = 20 \ \mbox{ns}$ single-qubit NOT gate, starting from the initial condition $|\psi(t=0)\rangle = |0\rangle$.  $P_2(t)$ has been multiplied by a factor of $20$ to illustrate its shape. (b) Dephasing-induced leakage for white noise.  The probability of leakage $\mathcal{P}_{\text{leakage}}^{(1)}$ as a function of gate time $T$ calculated using Eq.~(\ref{RabiL}) (solid curve) and by a corresponding master equation simulation (squares), with $T_{\varphi}^{(1)} = 100 \ \mu \mbox{s}$.  (c) Dephasing-induced leakage for low-frequency noise.  rom single-qubit NOT gate as a function of gate time $T$ for low-frequency noise.  The probability of leakage $\mathcal{P}_{\text{leakage}}^{(2)}$ as a function of gate time $T$ calculated using Eq.~(\ref{RabiL}) (solid curve) and by numerical integration of Eq.~(\ref{correlatedL}), with $T_{\varphi}^{(2)} = 1 \ \mu \mbox{s}$.}
\end{figure*}

\section{Single-Qubit NOT Gate}

Single-qubit operations are typically performed by applying a microwave drive resonant with the transition between the qubit states $|0\rangle$ and $|1\rangle$.  However, this drive can also cause off-resonant transitions from $|1\rangle$ to the auxiliary state $|2\rangle$.  In a rotating frame, this system is well-described by the Hamiltonan
\begin{equation}
\mathcal{H}_0 = \begin{pmatrix} 0 & \Omega/2 & 0 \\[0.1cm] \Omega^*/2 & 0 & \Omega/\sqrt{2} \\[0.1cm] 0 & \Omega^*/\sqrt{2} & -\eta \end{pmatrix},
\end{equation}
where $\Omega(t)$ is a (possibly complex) shaped microwave amplitude and $\eta = \omega_{01} - \omega_{12}$ is the transmon anharmonicity.  The dephasing Hamiltonian is taken to be $\mathcal{H}_{\text{noise}} = \varepsilon(t) V_0$ with
\begin{equation}
V_0 = \begin{pmatrix} 0 & 0 & 0 \\ 0 & 1 & 0 \\ 0 & 0 & 2\end{pmatrix},
\end{equation}
Proper pulse-shaping \cite{chow2010optimized,lucero2010reduced}, typically using a form known as DRAG \cite{motzoi2009simple}, allows the final unitary $U_0(T)$ (in the absence of dephasing) to have no coupling between the qubit and auxiliary states.  There will, however, be transient population, which will lead to leakage in the presence of dephasing.

\subsection{Perturbation Theory}

In what follows I will restrict to a real drive amplitude $\Omega(t)$, and use perturbation theory to calculate $U_0(t)$ for a single-qubit NOT gate.  To proceed, I split the Hamiltonian into $\mathcal{H}_0=\mathcal{H}_{0}^{(0)}+\mathcal{V}_0^{(0)}$, where
\begin{equation}
\mathcal{H}_0^{(0)} = \begin{pmatrix} 0 & \Omega/2 & 0 \\[0.1cm] \Omega/2 & 0 & 0 \\[0.1 cm] 0 & 0 & -\eta \end{pmatrix}
\end{equation}
and
\begin{equation}
\mathcal{V}_0^{(0)} = \begin{pmatrix} 0 & 0 & 0 \\[0.1cm] 0 & 0 & \Omega/\sqrt{2} \\[0.1 cm] 0 & \Omega/\sqrt{2} & 0 \end{pmatrix}.
\end{equation}
Similarly, I split the time-evolution operator into
\begin{equation}
U_0(t) = U_0^{(0)}(t) U_0^{(1)}(t),
\end{equation}
where 
\begin{align}
U_0^{(0)}(t) &= \mathcal{T} \exp \left[- i \int_0^t \mathcal{H}_{0}^{(0)}(t') dt' \right]  \nonumber \\
& = \begin{pmatrix} \cos(\theta/2) & -i \sin(\theta/2) & 0 \\[0.1cm] -i \sin(\theta/2) & \cos(\theta/2) & 0 \\[0.1cm] 0 & 0 & e^{i \eta t} \end{pmatrix}
\label{pertU0}
\end{align}
and
\begin{align}
U_0^{(1)}(t) & = \mathcal{T} \exp \left[- i \int_0^t \mathcal{H}_{0}^{(1)}(t') dt' \right] \nonumber \\
& \approx I - i \int_0^t \mathcal{H}_{0}^{(1)}(t')dt'
\end{align}
with
\begin{equation}
\mathcal{H}_{0}^{(1)} = \left[U_0^{(0)}(t) \right]^{\dagger} \mathcal{V}_{0}^{(0)} U_0^{(0)}(t).
\label{pertH1}
\end{equation}

Note that in Eq.~(\ref{pertU0}), we have suppressed the time-dependence of the rotation angle
\begin{equation}
\theta(t) = \int_0^{t} \Omega(t') dt'.
\end{equation}
A single-qubit NOT gate will result if $\theta(T) = \pi$.   Substituting Eq.~(\ref{pertU0}) in Eq.~(\ref{pertH1}) leads to
\begin{equation}
\mathcal{H}_{0}^{(1)} = \begin{pmatrix} 0 & 0 & i \alpha \\[0.1cm] 0 & 0 & \beta \\[0.1cm] -i \alpha^* & \beta^* & 0 \end{pmatrix},
\end{equation}
where 
\begin{align}
\alpha(t) &= \Omega e^{i \eta t} \sin(\theta/2) / \sqrt{2}, \nonumber \\
\beta(t)  &=  \Omega e^{i \eta t} \cos(\theta/2) / \sqrt{2}.
\end{align}
To calculate $U_{0}^{(1)}(t)$, one should integrate $\mathcal{H}_{0}^{(1)}$.  This will generally depend on the pulse shape $\Omega(t)$.  However, a useful approximation can be made by introducing an integration by parts and using the standard assumption that $\Omega(0) = \Omega(T) = 0$.   For example, consider the integral
\begin{align}
\int_0^t f(t') e^{i \eta t'} dt' & = \frac{1}{i \eta} \int_0^t f(t') \frac{d}{dt'} e^{i \eta t'} dt' \nonumber \\
& = \frac{1}{i \eta} f(t) e^{i \eta t} - \frac{1}{i \eta} \int_0^t f'(t') e^{i \eta t'} dt',
\label{integral_approx}
\end{align}
where we have assumed that $f(0) = 0$ in evaluating the first term.  The integration by parts can then be repeated for the second term, yielding a power series in $1/\eta$ and repeated derivatives of $f(t)$ (evaluated at the endpoints).  However, only the first term of Eq.~(\ref{integral_approx}) is needed here.  Using this approximation leads to
\begin{equation}
U_0^{(1)} \approx \begin{pmatrix} 1 & 0 & -i \alpha/\eta \\[0.1cm] 0 & 1 & -\beta/\eta \\[0.1cm] -i \alpha^*/\eta & \beta^*/\eta & 1 \end{pmatrix}.
\end{equation}
This expression is consistent with the adiabatic frame transformation used in \cite{motzoi2009simple}.


With these expressions for $U_0^{(0)}(t)$ and $U_0^{(1)}(t)$, one can calculate $V_1 = U_0^{\dagger} V_0 U_0 =  [U_{0}^{(0)} U_{0}^{(1)}]^{\dagger} V_0 U_{0}^{(0)} U_{0}^{(1)}$ to find
\begin{equation}
\mathcal{A}_{0 \to 2}(t) = \langle 2| V_1 (t) |0\rangle = - i \frac{\Omega}{\sqrt{2} \eta} e^{-i \eta t} \sin(\theta/2) 
\end{equation}
and
\begin{equation}
\mathcal{A}_{1 \to 2}(t) = \langle 2| V_1(t) |1\rangle = - \frac{\Omega}{\sqrt{2} \eta} e^{-i \eta t} \cos (\theta/2).
\end{equation}
Each of these oscillate with characteristic frequency $\eta$, so that the leakage, using Eq.~(\ref{generalL}) with $d_Q=2$ and $d_A = 1$, becomes
\begin{equation}
\mathcal{P}_{\text{leakage}} = \frac{1}{4 \eta^2} S(\eta) \int_{0}^T \Omega^2(t') dt'.
\end{equation}

For a pulse shape of the form 
\begin{equation}
\Omega(t) = \frac{\pi}{T} \left[ 1- \cos(2 \pi t/T)\right],
\end{equation}
this leakage can be calculated as
\begin{equation}
 \mathcal{P}_{\text{leakage}} = \frac{3 \pi^2}{8 \eta^2 T} S(\eta).
 \end{equation}
 For white noise, $S_0 = 2/T_{\varphi}^{(1)}$, this reduces to 
 \begin{equation}
 \mathcal{P}_{\text{leakage}}^{(1)} = \frac{3 \pi^2}{4 \eta^2 T T_{\varphi}^{(1)}}.
 \label{RabiL}
 \end{equation}
 For $1/f$ noise, with amplitude $S_1 = (1/ 2.6)  [T_{\varphi}^{(2)}]^{-2}$, the leakage evaluates to
 \begin{equation}
 \mathcal{P}_\text{leakage}^{(2)}  \approx \frac{\pi^2}{7 \eta^3 T [T_{\varphi}^{(2)}]^2}
 \label{RabiLlowf}
 \end{equation}
 
 \subsection{Comparison with DRAG}
 
For numerical simulations, I use a DRAG formulation $\Omega(t) = \Omega_x(t) - i \Omega_y(t)$ for a single-qubit NOT gate for a transmon with $\omega_{01}/2\pi = 6 \ \mbox{GHz}$ and $\eta/2\pi = 260 \ \mbox{MHz}$.  This optimization does not use the rotating-wave approximation, and allows for a drive frequency that is shifted from $\omega_{01}$ \cite{chen2014qubit}; more details are given in Appendix C.  The resulting state probabilities $P_0(t) = |\langle 0 | \psi(t) \rangle|^2$, $P_1(t) = |\langle 1 |\psi(t) \rangle|^2$, and $P_2(t) = |\langle 2 |\psi(t)\rangle$ for a $T = 20 \ \mbox{ns}$ gate are shown in Fig. 3(a), starting from the initial condition $|\psi(t=0)\rangle = |0\rangle$.  The leakage probability for white noise, calculated using Eq.~(\ref{RabiL}) with $T_{\varphi}^{(1)} = 100 \ \mu \mbox{s}$, is shown as a function of time $T$ in Fig. 3(b) (solid curve).  Also shown are results from a master equation simulation of the density matrix, after optimizing the pulse for each gate time. These two calculations are in excellent agreement, and exhibit typical values of order $10^{-6}$.   For comparison, the equivalent leakage probability quoted in \cite{google2023suppressing} is $5 \times 10^{-5}$ (as discussed in Sec. I of their Supplementary Information).  

Results for $1/f$ noise are shown in Fig.~3(c).  The leakage probability, calculated using Eq.~(\ref{RabiLlowf}) with $T_{\varphi}^{(2)} = 1 \ \mu \mbox{s}$, is shown as a function of time $T$ (solid curve).  Also shown are results from a numerical integration of Eq.~(\ref{correlatedL}), where the amplitudes $\mathcal{A}_{0 \to 2}(t)$ and $\mathcal{A}_{1 \to 2}(t)$ are found by numerical simulation of the Schr{\"o}dinger equation.  These two calculations are also in excellent agreement, and are much less than the leakage for white noise.
 
\section{Conclusion}

As discussed in the introduction, finding and removing leakage mechanisms remains an important task for quantum information processing with superconducting qubits.  In this paper I have analyzed how dephasing causes leakage in typical single- and two-qubit gates for transmon-style devices.  The calculated errors are approximately one order of magnitude smaller than those recently reported in \cite{google2023suppressing}, but may constitute an obstacle to continued progress.  A few observations are worth making.

As a first observation, the presence of nearby auxiliary energy states in transmons is due to their intrinsic nature as weakly anharmonic oscillators. Thus, a device with greater anharmonicity, such as fluxonium \cite{nguyen2019high}, would be an attractive alternative.  Fluxonium devices have the additional advantage of significantly reduced dephasing.  However, while single-qubit operations should be much less susceptible to leakage, two-qubit fluxonium gates that use excited states, such as recently those demonstrated in \cite{ficheux2021fast} and \cite{ding2023high}, may not be.  

This brings us to a second observation, namely that one could consider two-qubit operations that have reduced (transient) population in auxiliary states.  Examples include the iSWAP gate \cite{barends2019diabatic}, the dispersive gate demonstrated in \cite{jin2023versatile}, or the spectroscopic gate demonstrated in \cite{premaratne2019implementation}.  Combining such approaches with the fluxonium constitutes an important research direction.

As a final observation, one might consider relaxing the assumptions made in this paper.  The dominant one is the assumption that dephasing can be treated as a classical noise process $\varepsilon(t)$.  A full understanding of dephasing-induced leakage may require a more involved quantum treatment of the underlying superconducting circuit and its environment.  

\acknowledgments

I thank Williams College and the IQC at the University of Waterloo, where this work was initiated, for their support.  I also thank Ray Simmonds and Rob Schoelkopf for their encouragement.  This work was supported by the Army Research Office Grant \# W911NF-18-1-0056.  

\appendix

\section{Spectral Densities and Dephasing Times}

In this Appendix the conventions used in this paper to relate the spectral density of noise to relevant dephasing times are specified.  These relations largely follow Appendix B in O'Malley {\it et al.} \cite{omalley2015}, although with slight changes in notation and a factor of two (corresponding to the difference between the single-sided and double-sided spectral density conventions)  For noise spectrum $S(\omega)$, the dephasing factor $e^{-\mathcal{D}(t)}$ is found by taking the noise average of $\exp [-i \varphi(t) ]$, where 
\begin{equation}
\phi(t) = \int_{0}^t \varepsilon(t') dt',
\end{equation}
so that
\begin{align}
\mathcal{D}(t) &= \frac{1}{2} \langle \phi(t)^2 \rangle \nonumber \\
&= \frac{1}{2} \int_{0}^t dt' \int_{0}^t dt'' C(t'-t'') \nonumber \\
&= \frac{1}{\pi} \int_{-\infty}^{\infty} d\omega S(\omega) \left[ \frac{\sin(\omega t/2)}{\omega} \right]^2.
\end{align}
Note that while this assumes a Gaussian noise process, the perturbative calculations in the text can apply for non-Gaussian processes.

For white noise, $S(\omega) = S_0$, and one finds
\begin{align}
\mathcal{D}_1(t) &= \frac{1}{\pi} S_0 t \int_0^\infty dx \left[ \frac{\sin(x)}{x} \right]^2 \nonumber \\
&= \frac{1}{2} S_0 t \equiv t / T_{\varphi}^{(1)},
\end{align}
where $T_{\varphi}^{(1)}$ is the ``linear'' dephasing time.  Thus, for white noise acting on a single qubit, the spectral density $S_0 = 2 / T_{\varphi}^{(1)}$. 

For $1/f$ noise, $S(\omega) = S_1/|\omega|$, and one finds
\begin{equation}
\mathcal{D}_2(t) = \frac{S_1}{2 \pi} t^2 \int_{x_{\text{min}}}^{x_{\text{max}}} dx \left[ \frac{ \sin^2 x}{x^3} \right],
\end{equation}
where the limits of integration are given in terms of cutoff frequencies by $x_{\text{min}} = \pi f_{\text{min}} t$ and $x_{\text{max}} = \pi f_{\text{max}} t$.
The integral on the right-hand-side can be evaluated in terms of the cosine integral function $\mbox{Ci}(z)$, and the dominant behavior is given by the low frequency contribution so that
\begin{equation}
\mathcal{D}_2(t) \approx \frac{S_1}{2 \pi} t^2 \ln (c / x_{\text{min}}),
\end{equation}
where
\begin{equation}
\ln c = \frac{3}{2} - \gamma - \ln(2) \approx 0.23,
\end{equation}
with $\gamma \approx 0.577$ is Euler's constant.  The logarithmic factor is slowly varying, and with a lower-frequency cutoff of $f_{\mbox{min}} = 1 \ \mbox{Hz}$ and for timescales in the range of $20 \to 30 \ \mbox{ns}$ decreases from approximately $16.8$ to $16.4$.  Thus, one can approximate
\begin{equation}
\mathcal{D}_2(t) \approx S_1 2.6 t^2 \equiv \left[ t / T_{\varphi}^(2) \right]^2   
\end{equation}
That is, for a single-qubit ``quadratic'' dephasing time $T_{\varphi}^{(2)}$, the spectral density is approximately $S(\omega) = (1/2.6) [T_{\varphi}^{(2)}]^{-2} / |\omega|$.

The dephasing times $T_{\varphi}^{(1)} = 100 \ \mu \mbox{s}$ and $T_{\varphi}^{(2)} = 1 \ \mu \mbox{s}$ used in the text are chosen to provide identical errors for an ``idling'' duration of $T = 10 \ \mbox{ns}$.  In addition, the spectral densities used for the two-qubit gate in Sec. III are doubled to account for the dephasing of both qubits.

\section{Density Matrix Simulation for the Controlled-Phase Gate}

The two-qubit master equation simulations reported in Fig.2(b) are solutions of 
\begin{equation}
\frac{d \rho}{dt} = -i [H,\rho] + \lambda \left( L \rho L^{\dagger} - \frac{1}{2} L^{\dagger} L \rho - \frac{1}{2} \rho L^{\dagger} L \right)
\end{equation}
where
\begin{equation}
\rho = \begin{pmatrix} \rho_{00} & \rho_{01} \\ \rho_{10} & \rho_{11} \end{pmatrix},
\end{equation}
with initial condition $\rho_{00}(t=0) = 1$ (and all other components are zero).  The time-dependent Hamiltonian $H$ is given by Eq.~(\ref{phase_hamiltonian}) with $g/2\pi = 50 \ \mbox{MHz}$ and $\Delta(t) = 2 g / \tan[ \theta(t)]$, $\Delta(0)/2\pi = 1 \ \mbox{GHz}$, and $\theta(t)$ is given by  Eq~(\ref{theta_trajectory}).  The parameters $\theta_1$ and $\theta_2$ for $\theta(5)$ are numerically optimized for each gate time $T$ to produce a high-fidelity gate with net accumulated phase $\pi$.  

The Lindblad operator is given by
\begin{equation}
L = V_0 = \begin{pmatrix} 0 & 0 \\ 0 & 1 \end{pmatrix}.
 \end{equation}
 with constant rate $\lambda$.  This yields the dephasing evolution
\begin{equation}
\rho \to \begin{pmatrix} \rho_{00} & \rho_{01} e^{-\lambda t/2} \\ \rho_{10} e^{-\lambda t/2} & \rho_{11} \end{pmatrix}.
\end{equation}
The rate $\lambda = 4/T_{\varphi}^{(1)}$ is chosen to match the standard model of dephasing for states $|11\rangle$ and $|20\rangle$, whose coherences decay with rate $1/T_{\varphi,1} + 1/T_{\varphi_2} = 2/T_{\varphi}^{(1)}$, where $T_{\varphi,1}$ and $T_{\varphi,2}$ are the dephasing times for transmons $1$ and $2$, respectively.  

This master equation approach will agree with the noise-averaged time evolution due to $\mathcal{H}_{\text{noise}} = \varepsilon(t) V_0$ if
\begin{align}
e^{-\lambda t/2} &= \left\langle \exp \left[ -i \int_{0}^{t} \varepsilon(t') dt' \right] \right \rangle_{\text{noise}} \nonumber \\
& = \exp \left[ - \frac{1}{2} \int_{0}^{t} dt' \int_{0}^t dt'' C(t'-t'') \right]
\end{align}
This is true for the white noise spectrum with $C(t-t') = S_0 \delta(t-t')$ with $S_0 = \lambda = 4/T_{\varphi}^{(1)}$.  For rapid tuning, the master equation can be solved exactly, and one finds
\begin{equation}
\rho_{11}(T) = \frac{1}{2} - \frac{1}{2} e^{- \lambda T/4} \approx \frac{1}{8} \lambda T = \frac{1}{8} S_0 T = \frac{1}{2} \left( \frac{T}{T_{\varphi}^{(1)}} \right).
\end{equation}

\section{Density Matrix Simulation for the NOT Gate}
 The single-qubit master equation simulations reported in Fig.3(b) are solutions of
 \begin{equation}
\frac{d \rho}{dt} = -i [H,\rho] + \lambda \left( L \rho L^{\dagger} - \frac{1}{2} L^{\dagger} L \rho - \frac{1}{2} \rho L^{\dagger} L \right)
\end{equation}
for a four-level system
\begin{equation}
\rho = \begin{pmatrix} \rho_{00} & \rho_{01} & \rho_{02} & \rho_{03} \\ \rho_{10} & \rho_{11} & \rho_{12} & \rho_{13} \\ \rho_{20} & \rho_{21} & \rho_{22} & \rho_{23} \\ 
\rho_{30} & \rho_{31} & \rho_{32} & \rho_{33} \end{pmatrix},
\end{equation}
with initial condition $\rho_{00}(t=0) = 1$ (and all other components are zero).  The time-dependent Hamiltonian is $H = H_0 + H_{\text{drive}}$ where  
\begin{equation}
H_0 = \begin{pmatrix} 0 & 0 & 0 & 0 \\ 0 & \omega_{01} & 0 & 0 & 0 \\ 0 & 0 & 2 \omega_{01} - \eta & 0 \\ 0 & 0 & 0 & 3 \omega_{01} - \eta \end{pmatrix}
\end{equation}
with $\omega_{01}/2\pi = 6 \ \mbox{GHz}$ and $\eta/2\pi = 260 \ \mbox{MHz}$, and
\begin{equation}
H_{\text{drive}} = \left[ \Omega_x(t) \cos (\omega t) + \Omega_y(t) \sin(\omega t) \right] X
\end{equation}
with
\begin{equation}
X = \begin{pmatrix} 0 & 1 & 0 & 0 \\ 1 & 0 & \sqrt{2} & 0 \\ 0 & \sqrt{2} & 0 & \sqrt{3} \\ 0 & 0 & \sqrt{3} & 0 \end{pmatrix}.
\end{equation}
The pulse envelopes are of the form
\begin{align}
\Omega_x &= \Omega_1 [1 - \cos(2 \pi t/T)], \nonumber \\
\Omega_y &= \Omega_2 \sin(2 \pi t/T).
\end{align}
The parameters $\Omega_1$, $\Omega_2$, and $\omega$ are numerically optimized for each gate time $T$ to produce a high-fidelity NOT gate.

The Lindblad operator is
\begin{equation}
L = \begin{pmatrix} 0 & 0 & 0 & 0 \\ 0 & 1 & 0 & 0 \\ 0 & 0 & 2 & 0 \\ 0 & 0 & 0 & 3 \end{pmatrix},
\end{equation}
with rate $\lambda = 2/T_{\varphi}^{(1)}$.

\bibliography{dephasingbib}

\begin{thebibliography}{33}%
\makeatletter
\providecommand \@ifxundefined [1]{%
 \@ifx{#1\undefined}
}%
\providecommand \@ifnum [1]{%
 \ifnum #1\expandafter \@firstoftwo
 \else \expandafter \@secondoftwo
 \fi
}%
\providecommand \@ifx [1]{%
 \ifx #1\expandafter \@firstoftwo
 \else \expandafter \@secondoftwo
 \fi
}%
\providecommand \natexlab [1]{#1}%
\providecommand \enquote  [1]{``#1''}%
\providecommand \bibnamefont  [1]{#1}%
\providecommand \bibfnamefont [1]{#1}%
\providecommand \citenamefont [1]{#1}%
\providecommand \href@noop [0]{\@secondoftwo}%
\providecommand \href [0]{\begingroup \@sanitize@url \@href}%
\providecommand \@href[1]{\@@startlink{#1}\@@href}%
\providecommand \@@href[1]{\endgroup#1\@@endlink}%
\providecommand \@sanitize@url [0]{\catcode `\\12\catcode `\$12\catcode
  `\&12\catcode `\#12\catcode `\^12\catcode `\_12\catcode `\%12\relax}%
\providecommand \@@startlink[1]{}%
\providecommand \@@endlink[0]{}%
\providecommand \url  [0]{\begingroup\@sanitize@url \@url }%
\providecommand \@url [1]{\endgroup\@href {#1}{\urlprefix }}%
\providecommand \urlprefix  [0]{URL }%
\providecommand \Eprint [0]{\href }%
\providecommand \doibase [0]{https://doi.org/}%
\providecommand \selectlanguage [0]{\@gobble}%
\providecommand \bibinfo  [0]{\@secondoftwo}%
\providecommand \bibfield  [0]{\@secondoftwo}%
\providecommand \translation [1]{[#1]}%
\providecommand \BibitemOpen [0]{}%
\providecommand \bibitemStop [0]{}%
\providecommand \bibitemNoStop [0]{.\EOS\space}%
\providecommand \EOS [0]{\spacefactor3000\relax}%
\providecommand \BibitemShut  [1]{\csname bibitem#1\endcsname}%
\let\auto@bib@innerbib\@empty
\bibitem [{\citenamefont {Martinis}\ \emph {et~al.}(2002)\citenamefont
  {Martinis}, \citenamefont {Nam}, \citenamefont {Aumentado},\ and\
  \citenamefont {Urbina}}]{martinis2002rabi}%
  \BibitemOpen
  \bibfield  {author} {\bibinfo {author} {\bibfnamefont {J.~M.}\ \bibnamefont
  {Martinis}}, \bibinfo {author} {\bibfnamefont {S.}~\bibnamefont {Nam}},
  \bibinfo {author} {\bibfnamefont {J.}~\bibnamefont {Aumentado}},\ and\
  \bibinfo {author} {\bibfnamefont {C.}~\bibnamefont {Urbina}},\ }\bibfield
  {title} {\bibinfo {title} {Rabi oscillations in a large josephson-junction
  qubit},\ }\href@noop {} {\bibfield  {journal} {\bibinfo  {journal} {Physical
  review letters}\ }\textbf {\bibinfo {volume} {89}},\ \bibinfo {pages}
  {117901} (\bibinfo {year} {2002})}\BibitemShut {NoStop}%
\bibitem [{\citenamefont {Vion}\ \emph {et~al.}(2002)\citenamefont {Vion},
  \citenamefont {Aassime}, \citenamefont {Cottet}, \citenamefont {Joyez},
  \citenamefont {Pothier}, \citenamefont {Urbina}, \citenamefont {Esteve},\
  and\ \citenamefont {Devoret}}]{vion2002manipulating}%
  \BibitemOpen
  \bibfield  {author} {\bibinfo {author} {\bibfnamefont {D.}~\bibnamefont
  {Vion}}, \bibinfo {author} {\bibfnamefont {A.}~\bibnamefont {Aassime}},
  \bibinfo {author} {\bibfnamefont {A.}~\bibnamefont {Cottet}}, \bibinfo
  {author} {\bibfnamefont {P.}~\bibnamefont {Joyez}}, \bibinfo {author}
  {\bibfnamefont {H.}~\bibnamefont {Pothier}}, \bibinfo {author} {\bibfnamefont
  {C.}~\bibnamefont {Urbina}}, \bibinfo {author} {\bibfnamefont
  {D.}~\bibnamefont {Esteve}},\ and\ \bibinfo {author} {\bibfnamefont {M.~H.}\
  \bibnamefont {Devoret}},\ }\bibfield  {title} {\bibinfo {title} {Manipulating
  the quantum state of an electrical circuit},\ }\href@noop {} {\bibfield
  {journal} {\bibinfo  {journal} {Science}\ }\textbf {\bibinfo {volume}
  {296}},\ \bibinfo {pages} {886} (\bibinfo {year} {2002})}\BibitemShut
  {NoStop}%
\bibitem [{\citenamefont {Chiorescu}\ \emph {et~al.}(2003)\citenamefont
  {Chiorescu}, \citenamefont {Nakamura}, \citenamefont {Harmans},\ and\
  \citenamefont {Mooij}}]{chiorescu2003coherent}%
  \BibitemOpen
  \bibfield  {author} {\bibinfo {author} {\bibfnamefont {I.}~\bibnamefont
  {Chiorescu}}, \bibinfo {author} {\bibfnamefont {Y.}~\bibnamefont {Nakamura}},
  \bibinfo {author} {\bibfnamefont {C.~M.}\ \bibnamefont {Harmans}},\ and\
  \bibinfo {author} {\bibfnamefont {J.}~\bibnamefont {Mooij}},\ }\bibfield
  {title} {\bibinfo {title} {Coherent quantum dynamics of a superconducting
  flux qubit},\ }\href@noop {} {\bibfield  {journal} {\bibinfo  {journal}
  {Science}\ }\textbf {\bibinfo {volume} {299}},\ \bibinfo {pages} {1869}
  (\bibinfo {year} {2003})}\BibitemShut {NoStop}%
\bibitem [{\citenamefont {Pashkin}\ \emph {et~al.}(2003)\citenamefont
  {Pashkin}, \citenamefont {Yamamoto}, \citenamefont {Astafiev}, \citenamefont
  {Nakamura}, \citenamefont {Averin},\ and\ \citenamefont
  {Tsai}}]{pashkin2003quantum}%
  \BibitemOpen
  \bibfield  {author} {\bibinfo {author} {\bibfnamefont {Y.~A.}\ \bibnamefont
  {Pashkin}}, \bibinfo {author} {\bibfnamefont {T.}~\bibnamefont {Yamamoto}},
  \bibinfo {author} {\bibfnamefont {O.}~\bibnamefont {Astafiev}}, \bibinfo
  {author} {\bibfnamefont {Y.}~\bibnamefont {Nakamura}}, \bibinfo {author}
  {\bibfnamefont {D.}~\bibnamefont {Averin}},\ and\ \bibinfo {author}
  {\bibfnamefont {J.}~\bibnamefont {Tsai}},\ }\bibfield  {title} {\bibinfo
  {title} {Quantum oscillations in two coupled charge qubits},\ }\href@noop {}
  {\bibfield  {journal} {\bibinfo  {journal} {Nature}\ }\textbf {\bibinfo
  {volume} {421}},\ \bibinfo {pages} {823} (\bibinfo {year}
  {2003})}\BibitemShut {NoStop}%
\bibitem [{\citenamefont {Berkley}\ \emph {et~al.}(2003)\citenamefont
  {Berkley}, \citenamefont {Xu}, \citenamefont {Ramos}, \citenamefont {Gubrud},
  \citenamefont {Strauch}, \citenamefont {Johnson}, \citenamefont {Anderson},
  \citenamefont {Dragt}, \citenamefont {Lobb},\ and\ \citenamefont
  {Wellstood}}]{berkley2003entangled}%
  \BibitemOpen
  \bibfield  {author} {\bibinfo {author} {\bibfnamefont {A.}~\bibnamefont
  {Berkley}}, \bibinfo {author} {\bibfnamefont {H.}~\bibnamefont {Xu}},
  \bibinfo {author} {\bibfnamefont {R.}~\bibnamefont {Ramos}}, \bibinfo
  {author} {\bibfnamefont {M.}~\bibnamefont {Gubrud}}, \bibinfo {author}
  {\bibfnamefont {F.}~\bibnamefont {Strauch}}, \bibinfo {author} {\bibfnamefont
  {P.}~\bibnamefont {Johnson}}, \bibinfo {author} {\bibfnamefont
  {J.}~\bibnamefont {Anderson}}, \bibinfo {author} {\bibfnamefont
  {A.}~\bibnamefont {Dragt}}, \bibinfo {author} {\bibfnamefont
  {C.}~\bibnamefont {Lobb}},\ and\ \bibinfo {author} {\bibfnamefont
  {F.}~\bibnamefont {Wellstood}},\ }\bibfield  {title} {\bibinfo {title}
  {Entangled macroscopic quantum states in two superconducting qubits},\
  }\href@noop {} {\bibfield  {journal} {\bibinfo  {journal} {Science}\ }\textbf
  {\bibinfo {volume} {300}},\ \bibinfo {pages} {1548} (\bibinfo {year}
  {2003})}\BibitemShut {NoStop}%
\bibitem [{\citenamefont {Yamamoto}\ \emph {et~al.}(2003)\citenamefont
  {Yamamoto}, \citenamefont {Pashkin}, \citenamefont {Astafiev}, \citenamefont
  {Nakamura},\ and\ \citenamefont {Tsai}}]{yamamoto2003demonstration}%
  \BibitemOpen
  \bibfield  {author} {\bibinfo {author} {\bibfnamefont {T.}~\bibnamefont
  {Yamamoto}}, \bibinfo {author} {\bibfnamefont {Y.~A.}\ \bibnamefont
  {Pashkin}}, \bibinfo {author} {\bibfnamefont {O.}~\bibnamefont {Astafiev}},
  \bibinfo {author} {\bibfnamefont {Y.}~\bibnamefont {Nakamura}},\ and\
  \bibinfo {author} {\bibfnamefont {J.-S.}\ \bibnamefont {Tsai}},\ }\bibfield
  {title} {\bibinfo {title} {Demonstration of conditional gate operation using
  superconducting charge qubits},\ }\href@noop {} {\bibfield  {journal}
  {\bibinfo  {journal} {Nature}\ }\textbf {\bibinfo {volume} {425}},\ \bibinfo
  {pages} {941} (\bibinfo {year} {2003})}\BibitemShut {NoStop}%
\bibitem [{\citenamefont {Gupta}\ \emph {et~al.}(2024)\citenamefont {Gupta},
  \citenamefont {Sundaresan}, \citenamefont {Alexander}, \citenamefont {Wood},
  \citenamefont {Merkel}, \citenamefont {Healy}, \citenamefont {Hillenbrand},
  \citenamefont {Jochym-O’Connor}, \citenamefont {Wootton}, \citenamefont
  {Yoder} \emph {et~al.}}]{gupta2024encoding}%
  \BibitemOpen
  \bibfield  {author} {\bibinfo {author} {\bibfnamefont {R.~S.}\ \bibnamefont
  {Gupta}}, \bibinfo {author} {\bibfnamefont {N.}~\bibnamefont {Sundaresan}},
  \bibinfo {author} {\bibfnamefont {T.}~\bibnamefont {Alexander}}, \bibinfo
  {author} {\bibfnamefont {C.~J.}\ \bibnamefont {Wood}}, \bibinfo {author}
  {\bibfnamefont {S.~T.}\ \bibnamefont {Merkel}}, \bibinfo {author}
  {\bibfnamefont {M.~B.}\ \bibnamefont {Healy}}, \bibinfo {author}
  {\bibfnamefont {M.}~\bibnamefont {Hillenbrand}}, \bibinfo {author}
  {\bibfnamefont {T.}~\bibnamefont {Jochym-O’Connor}}, \bibinfo {author}
  {\bibfnamefont {J.~R.}\ \bibnamefont {Wootton}}, \bibinfo {author}
  {\bibfnamefont {T.~J.}\ \bibnamefont {Yoder}}, \emph {et~al.},\ }\bibfield
  {title} {\bibinfo {title} {Encoding a magic state with beyond break-even
  fidelity},\ }\href@noop {} {\bibfield  {journal} {\bibinfo  {journal}
  {Nature}\ }\textbf {\bibinfo {volume} {625}},\ \bibinfo {pages} {259}
  (\bibinfo {year} {2024})}\BibitemShut {NoStop}%
\bibitem [{\citenamefont {{Google~Quantum~AI}}()}]{google2024qec}%
  \BibitemOpen
  \bibfield  {author} {\bibinfo {author} {\bibnamefont {{Google~Quantum~AI}}},\
  }\bibfield  {title} {\bibinfo {title} {Suppressing quantum errors by scaling
  a surface code logical qubit},\ }\href@noop {} {\bibinfo  {journal} {Nature}\
  }\BibitemShut {NoStop}%
\bibitem [{\citenamefont {{Google~Quantum~AI}}(2023)}]{google2023suppressing}%
  \BibitemOpen
\bibfield  {journal} {  }\bibfield  {author} {\bibinfo {author} {\bibnamefont
  {{Google~Quantum~AI}}},\ }\bibfield  {title} {\bibinfo {title} {Suppressing
  quantum errors by scaling a surface code logical qubit},\ }\href@noop {}
  {\bibfield  {journal} {\bibinfo  {journal} {Nature}\ }\textbf {\bibinfo
  {volume} {614}},\ \bibinfo {pages} {676} (\bibinfo {year}
  {2023})}\BibitemShut {NoStop}%
\bibitem [{\citenamefont {Chen}\ \emph {et~al.}(2016)\citenamefont {Chen},
  \citenamefont {Kelly}, \citenamefont {Quintana}, \citenamefont {Barends},
  \citenamefont {Campbell}, \citenamefont {Chen}, \citenamefont {Chiaro},
  \citenamefont {Dunsworth}, \citenamefont {Fowler}, \citenamefont {Lucero}
  \emph {et~al.}}]{chen2016measuring}%
  \BibitemOpen
  \bibfield  {author} {\bibinfo {author} {\bibfnamefont {Z.}~\bibnamefont
  {Chen}}, \bibinfo {author} {\bibfnamefont {J.}~\bibnamefont {Kelly}},
  \bibinfo {author} {\bibfnamefont {C.}~\bibnamefont {Quintana}}, \bibinfo
  {author} {\bibfnamefont {R.}~\bibnamefont {Barends}}, \bibinfo {author}
  {\bibfnamefont {B.}~\bibnamefont {Campbell}}, \bibinfo {author}
  {\bibfnamefont {Y.}~\bibnamefont {Chen}}, \bibinfo {author} {\bibfnamefont
  {B.}~\bibnamefont {Chiaro}}, \bibinfo {author} {\bibfnamefont
  {A.}~\bibnamefont {Dunsworth}}, \bibinfo {author} {\bibfnamefont
  {A.}~\bibnamefont {Fowler}}, \bibinfo {author} {\bibfnamefont
  {E.}~\bibnamefont {Lucero}}, \emph {et~al.},\ }\bibfield  {title} {\bibinfo
  {title} {Measuring and suppressing quantum state leakage in a superconducting
  qubit},\ }\href@noop {} {\bibfield  {journal} {\bibinfo  {journal} {Physical
  review letters}\ }\textbf {\bibinfo {volume} {116}},\ \bibinfo {pages}
  {020501} (\bibinfo {year} {2016})}\BibitemShut {NoStop}%
\bibitem [{\citenamefont {Barends}\ \emph {et~al.}(2014)\citenamefont
  {Barends}, \citenamefont {Kelly}, \citenamefont {Megrant}, \citenamefont
  {Veitia}, \citenamefont {Sank}, \citenamefont {Jeffrey}, \citenamefont
  {White}, \citenamefont {Mutus}, \citenamefont {Fowler}, \citenamefont
  {Campbell} \emph {et~al.}}]{barends2014superconducting}%
  \BibitemOpen
  \bibfield  {author} {\bibinfo {author} {\bibfnamefont {R.}~\bibnamefont
  {Barends}}, \bibinfo {author} {\bibfnamefont {J.}~\bibnamefont {Kelly}},
  \bibinfo {author} {\bibfnamefont {A.}~\bibnamefont {Megrant}}, \bibinfo
  {author} {\bibfnamefont {A.}~\bibnamefont {Veitia}}, \bibinfo {author}
  {\bibfnamefont {D.}~\bibnamefont {Sank}}, \bibinfo {author} {\bibfnamefont
  {E.}~\bibnamefont {Jeffrey}}, \bibinfo {author} {\bibfnamefont {T.~C.}\
  \bibnamefont {White}}, \bibinfo {author} {\bibfnamefont {J.}~\bibnamefont
  {Mutus}}, \bibinfo {author} {\bibfnamefont {A.~G.}\ \bibnamefont {Fowler}},
  \bibinfo {author} {\bibfnamefont {B.}~\bibnamefont {Campbell}}, \emph
  {et~al.},\ }\bibfield  {title} {\bibinfo {title} {Superconducting quantum
  circuits at the surface code threshold for fault tolerance},\ }\href@noop {}
  {\bibfield  {journal} {\bibinfo  {journal} {Nature}\ }\textbf {\bibinfo
  {volume} {508}},\ \bibinfo {pages} {500} (\bibinfo {year}
  {2014})}\BibitemShut {NoStop}%
\bibitem [{\citenamefont {Kelly}\ \emph {et~al.}(2014)\citenamefont {Kelly},
  \citenamefont {Barends}, \citenamefont {Campbell}, \citenamefont {Chen},
  \citenamefont {Chen}, \citenamefont {Chiaro}, \citenamefont {Dunsworth},
  \citenamefont {Fowler}, \citenamefont {Hoi}, \citenamefont {Jeffrey} \emph
  {et~al.}}]{kelly2014optimal}%
  \BibitemOpen
  \bibfield  {author} {\bibinfo {author} {\bibfnamefont {J.}~\bibnamefont
  {Kelly}}, \bibinfo {author} {\bibfnamefont {R.}~\bibnamefont {Barends}},
  \bibinfo {author} {\bibfnamefont {B.}~\bibnamefont {Campbell}}, \bibinfo
  {author} {\bibfnamefont {Y.}~\bibnamefont {Chen}}, \bibinfo {author}
  {\bibfnamefont {Z.}~\bibnamefont {Chen}}, \bibinfo {author} {\bibfnamefont
  {B.}~\bibnamefont {Chiaro}}, \bibinfo {author} {\bibfnamefont
  {A.}~\bibnamefont {Dunsworth}}, \bibinfo {author} {\bibfnamefont {A.~G.}\
  \bibnamefont {Fowler}}, \bibinfo {author} {\bibfnamefont {I.-C.}\
  \bibnamefont {Hoi}}, \bibinfo {author} {\bibfnamefont {E.}~\bibnamefont
  {Jeffrey}}, \emph {et~al.},\ }\bibfield  {title} {\bibinfo {title} {Optimal
  quantum control using randomized benchmarking},\ }\href@noop {} {\bibfield
  {journal} {\bibinfo  {journal} {Physical review letters}\ }\textbf {\bibinfo
  {volume} {112}},\ \bibinfo {pages} {240504} (\bibinfo {year}
  {2014})}\BibitemShut {NoStop}%
\bibitem [{\citenamefont {Miao}\ \emph {et~al.}(2023)\citenamefont {Miao},
  \citenamefont {McEwen}, \citenamefont {Atalaya}, \citenamefont {Kafri},
  \citenamefont {Pryadko}, \citenamefont {Bengtsson}, \citenamefont {Opremcak},
  \citenamefont {Satzinger}, \citenamefont {Chen}, \citenamefont {Klimov} \emph
  {et~al.}}]{miao2023overcoming}%
  \BibitemOpen
  \bibfield  {author} {\bibinfo {author} {\bibfnamefont {K.~C.}\ \bibnamefont
  {Miao}}, \bibinfo {author} {\bibfnamefont {M.}~\bibnamefont {McEwen}},
  \bibinfo {author} {\bibfnamefont {J.}~\bibnamefont {Atalaya}}, \bibinfo
  {author} {\bibfnamefont {D.}~\bibnamefont {Kafri}}, \bibinfo {author}
  {\bibfnamefont {L.~P.}\ \bibnamefont {Pryadko}}, \bibinfo {author}
  {\bibfnamefont {A.}~\bibnamefont {Bengtsson}}, \bibinfo {author}
  {\bibfnamefont {A.}~\bibnamefont {Opremcak}}, \bibinfo {author}
  {\bibfnamefont {K.~J.}\ \bibnamefont {Satzinger}}, \bibinfo {author}
  {\bibfnamefont {Z.}~\bibnamefont {Chen}}, \bibinfo {author} {\bibfnamefont
  {P.~V.}\ \bibnamefont {Klimov}}, \emph {et~al.},\ }\bibfield  {title}
  {\bibinfo {title} {Overcoming leakage in quantum error correction},\
  }\href@noop {} {\bibfield  {journal} {\bibinfo  {journal} {Nature Physics}\
  }\textbf {\bibinfo {volume} {19}},\ \bibinfo {pages} {1780} (\bibinfo {year}
  {2023})}\BibitemShut {NoStop}%
\bibitem [{\citenamefont {Strauch}\ \emph {et~al.}(2003)\citenamefont
  {Strauch}, \citenamefont {Johnson}, \citenamefont {Dragt}, \citenamefont
  {Lobb}, \citenamefont {Anderson},\ and\ \citenamefont
  {Wellstood}}]{strauch2003quantum}%
  \BibitemOpen
  \bibfield  {author} {\bibinfo {author} {\bibfnamefont {F.~W.}\ \bibnamefont
  {Strauch}}, \bibinfo {author} {\bibfnamefont {P.~R.}\ \bibnamefont
  {Johnson}}, \bibinfo {author} {\bibfnamefont {A.~J.}\ \bibnamefont {Dragt}},
  \bibinfo {author} {\bibfnamefont {C.}~\bibnamefont {Lobb}}, \bibinfo {author}
  {\bibfnamefont {J.}~\bibnamefont {Anderson}},\ and\ \bibinfo {author}
  {\bibfnamefont {F.}~\bibnamefont {Wellstood}},\ }\bibfield  {title} {\bibinfo
  {title} {Quantum logic gates for coupled superconducting phase qubits},\
  }\href@noop {} {\bibfield  {journal} {\bibinfo  {journal} {Physical review
  letters}\ }\textbf {\bibinfo {volume} {91}},\ \bibinfo {pages} {167005}
  (\bibinfo {year} {2003})}\BibitemShut {NoStop}%
\bibitem [{\citenamefont {DiCarlo}\ \emph {et~al.}(2009)\citenamefont
  {DiCarlo}, \citenamefont {Chow}, \citenamefont {Gambetta}, \citenamefont
  {Bishop}, \citenamefont {Johnson}, \citenamefont {Schuster}, \citenamefont
  {Majer}, \citenamefont {Blais}, \citenamefont {Frunzio}, \citenamefont
  {Girvin} \emph {et~al.}}]{dicarlo2009demonstration}%
  \BibitemOpen
  \bibfield  {author} {\bibinfo {author} {\bibfnamefont {L.}~\bibnamefont
  {DiCarlo}}, \bibinfo {author} {\bibfnamefont {J.~M.}\ \bibnamefont {Chow}},
  \bibinfo {author} {\bibfnamefont {J.~M.}\ \bibnamefont {Gambetta}}, \bibinfo
  {author} {\bibfnamefont {L.~S.}\ \bibnamefont {Bishop}}, \bibinfo {author}
  {\bibfnamefont {B.~R.}\ \bibnamefont {Johnson}}, \bibinfo {author}
  {\bibfnamefont {D.}~\bibnamefont {Schuster}}, \bibinfo {author}
  {\bibfnamefont {J.}~\bibnamefont {Majer}}, \bibinfo {author} {\bibfnamefont
  {A.}~\bibnamefont {Blais}}, \bibinfo {author} {\bibfnamefont
  {L.}~\bibnamefont {Frunzio}}, \bibinfo {author} {\bibfnamefont
  {S.}~\bibnamefont {Girvin}}, \emph {et~al.},\ }\bibfield  {title} {\bibinfo
  {title} {Demonstration of two-qubit algorithms with a superconducting quantum
  processor},\ }\href@noop {} {\bibfield  {journal} {\bibinfo  {journal}
  {Nature}\ }\textbf {\bibinfo {volume} {460}},\ \bibinfo {pages} {240}
  (\bibinfo {year} {2009})}\BibitemShut {NoStop}%
\bibitem [{\citenamefont {Rol}\ \emph {et~al.}(2019)\citenamefont {Rol},
  \citenamefont {Battistel}, \citenamefont {Malinowski}, \citenamefont
  {Bultink}, \citenamefont {Tarasinski}, \citenamefont {Vollmer}, \citenamefont
  {Haider}, \citenamefont {Muthusubramanian}, \citenamefont {Bruno},
  \citenamefont {Terhal} \emph {et~al.}}]{rol2019fast}%
  \BibitemOpen
  \bibfield  {author} {\bibinfo {author} {\bibfnamefont {M.}~\bibnamefont
  {Rol}}, \bibinfo {author} {\bibfnamefont {F.}~\bibnamefont {Battistel}},
  \bibinfo {author} {\bibfnamefont {F.}~\bibnamefont {Malinowski}}, \bibinfo
  {author} {\bibfnamefont {C.}~\bibnamefont {Bultink}}, \bibinfo {author}
  {\bibfnamefont {B.}~\bibnamefont {Tarasinski}}, \bibinfo {author}
  {\bibfnamefont {R.}~\bibnamefont {Vollmer}}, \bibinfo {author} {\bibfnamefont
  {N.}~\bibnamefont {Haider}}, \bibinfo {author} {\bibfnamefont
  {N.}~\bibnamefont {Muthusubramanian}}, \bibinfo {author} {\bibfnamefont
  {A.}~\bibnamefont {Bruno}}, \bibinfo {author} {\bibfnamefont
  {B.}~\bibnamefont {Terhal}}, \emph {et~al.},\ }\bibfield  {title} {\bibinfo
  {title} {Fast, high-fidelity conditional-phase gate exploiting leakage
  interference in weakly anharmonic superconducting qubits},\ }\href@noop {}
  {\bibfield  {journal} {\bibinfo  {journal} {Physical review letters}\
  }\textbf {\bibinfo {volume} {123}},\ \bibinfo {pages} {120502} (\bibinfo
  {year} {2019})}\BibitemShut {NoStop}%
\bibitem [{\citenamefont {Neg{\^\i}rneac}\ \emph {et~al.}(2021)\citenamefont
  {Neg{\^\i}rneac}, \citenamefont {Ali}, \citenamefont {Muthusubramanian},
  \citenamefont {Battistel}, \citenamefont {Sagastizabal}, \citenamefont
  {Moreira}, \citenamefont {Marques}, \citenamefont {Vlothuizen}, \citenamefont
  {Beekman}, \citenamefont {Zachariadis} \emph {et~al.}}]{negirneac2021high}%
  \BibitemOpen
  \bibfield  {author} {\bibinfo {author} {\bibfnamefont {V.}~\bibnamefont
  {Neg{\^\i}rneac}}, \bibinfo {author} {\bibfnamefont {H.}~\bibnamefont {Ali}},
  \bibinfo {author} {\bibfnamefont {N.}~\bibnamefont {Muthusubramanian}},
  \bibinfo {author} {\bibfnamefont {F.}~\bibnamefont {Battistel}}, \bibinfo
  {author} {\bibfnamefont {R.}~\bibnamefont {Sagastizabal}}, \bibinfo {author}
  {\bibfnamefont {M.}~\bibnamefont {Moreira}}, \bibinfo {author} {\bibfnamefont
  {J.}~\bibnamefont {Marques}}, \bibinfo {author} {\bibfnamefont
  {W.}~\bibnamefont {Vlothuizen}}, \bibinfo {author} {\bibfnamefont
  {M.}~\bibnamefont {Beekman}}, \bibinfo {author} {\bibfnamefont
  {C.}~\bibnamefont {Zachariadis}}, \emph {et~al.},\ }\bibfield  {title}
  {\bibinfo {title} {High-fidelity controlled-z gate with maximal intermediate
  leakage operating at the speed limit in a superconducting quantum
  processor},\ }\href@noop {} {\bibfield  {journal} {\bibinfo  {journal}
  {Physical Review Letters}\ }\textbf {\bibinfo {volume} {126}},\ \bibinfo
  {pages} {220502} (\bibinfo {year} {2021})}\BibitemShut {NoStop}%
\bibitem [{\citenamefont {Chen}\ \emph {et~al.}(2014)\citenamefont {Chen},
  \citenamefont {Neill}, \citenamefont {Roushan}, \citenamefont {Leung},
  \citenamefont {Fang}, \citenamefont {Barends}, \citenamefont {Kelly},
  \citenamefont {Campbell}, \citenamefont {Chen}, \citenamefont {Chiaro} \emph
  {et~al.}}]{chen2014qubit}%
  \BibitemOpen
  \bibfield  {author} {\bibinfo {author} {\bibfnamefont {Y.}~\bibnamefont
  {Chen}}, \bibinfo {author} {\bibfnamefont {C.}~\bibnamefont {Neill}},
  \bibinfo {author} {\bibfnamefont {P.}~\bibnamefont {Roushan}}, \bibinfo
  {author} {\bibfnamefont {N.}~\bibnamefont {Leung}}, \bibinfo {author}
  {\bibfnamefont {M.}~\bibnamefont {Fang}}, \bibinfo {author} {\bibfnamefont
  {R.}~\bibnamefont {Barends}}, \bibinfo {author} {\bibfnamefont
  {J.}~\bibnamefont {Kelly}}, \bibinfo {author} {\bibfnamefont
  {B.}~\bibnamefont {Campbell}}, \bibinfo {author} {\bibfnamefont
  {Z.}~\bibnamefont {Chen}}, \bibinfo {author} {\bibfnamefont {B.}~\bibnamefont
  {Chiaro}}, \emph {et~al.},\ }\bibfield  {title} {\bibinfo {title} {Qubit
  architecture with high coherence and fast tunable coupling},\ }\href@noop {}
  {\bibfield  {journal} {\bibinfo  {journal} {Physical review letters}\
  }\textbf {\bibinfo {volume} {113}},\ \bibinfo {pages} {220502} (\bibinfo
  {year} {2014})}\BibitemShut {NoStop}%
\bibitem [{\citenamefont {Ganzhorn}\ \emph {et~al.}(2020)\citenamefont
  {Ganzhorn}, \citenamefont {Salis}, \citenamefont {Egger}, \citenamefont
  {Fuhrer}, \citenamefont {Mergenthaler}, \citenamefont {M{\"u}ller},
  \citenamefont {M{\"u}ller}, \citenamefont {Paredes}, \citenamefont {Pechal},
  \citenamefont {Werninghaus} \emph {et~al.}}]{ganzhorn2020benchmarking}%
  \BibitemOpen
  \bibfield  {author} {\bibinfo {author} {\bibfnamefont {M.}~\bibnamefont
  {Ganzhorn}}, \bibinfo {author} {\bibfnamefont {G.}~\bibnamefont {Salis}},
  \bibinfo {author} {\bibfnamefont {D.}~\bibnamefont {Egger}}, \bibinfo
  {author} {\bibfnamefont {A.}~\bibnamefont {Fuhrer}}, \bibinfo {author}
  {\bibfnamefont {M.}~\bibnamefont {Mergenthaler}}, \bibinfo {author}
  {\bibfnamefont {C.}~\bibnamefont {M{\"u}ller}}, \bibinfo {author}
  {\bibfnamefont {P.}~\bibnamefont {M{\"u}ller}}, \bibinfo {author}
  {\bibfnamefont {S.}~\bibnamefont {Paredes}}, \bibinfo {author} {\bibfnamefont
  {M.}~\bibnamefont {Pechal}}, \bibinfo {author} {\bibfnamefont
  {M.}~\bibnamefont {Werninghaus}}, \emph {et~al.},\ }\bibfield  {title}
  {\bibinfo {title} {Benchmarking the noise sensitivity of different parametric
  two-qubit gates in a single superconducting quantum computing platform},\
  }\href@noop {} {\bibfield  {journal} {\bibinfo  {journal} {Physical Review
  Research}\ }\textbf {\bibinfo {volume} {2}},\ \bibinfo {pages} {033447}
  (\bibinfo {year} {2020})}\BibitemShut {NoStop}%
\bibitem [{\citenamefont {Jin}\ \emph {et~al.}(2023)\citenamefont {Jin},
  \citenamefont {Cicak}, \citenamefont {Parrott}, \citenamefont {Kotler},
  \citenamefont {Lecocq}, \citenamefont {Teufel}, \citenamefont {Aumentado},
  \citenamefont {Kapit},\ and\ \citenamefont {Simmonds}}]{jin2023versatile}%
  \BibitemOpen
  \bibfield  {author} {\bibinfo {author} {\bibfnamefont {X.}~\bibnamefont
  {Jin}}, \bibinfo {author} {\bibfnamefont {K.}~\bibnamefont {Cicak}}, \bibinfo
  {author} {\bibfnamefont {Z.}~\bibnamefont {Parrott}}, \bibinfo {author}
  {\bibfnamefont {S.}~\bibnamefont {Kotler}}, \bibinfo {author} {\bibfnamefont
  {F.}~\bibnamefont {Lecocq}}, \bibinfo {author} {\bibfnamefont
  {J.}~\bibnamefont {Teufel}}, \bibinfo {author} {\bibfnamefont
  {J.}~\bibnamefont {Aumentado}}, \bibinfo {author} {\bibfnamefont
  {E.}~\bibnamefont {Kapit}},\ and\ \bibinfo {author} {\bibfnamefont
  {R.}~\bibnamefont {Simmonds}},\ }\bibfield  {title} {\bibinfo {title}
  {Versatile parametric coupling between two statically decoupled transmon
  qubits},\ }\href@noop {} {\bibfield  {journal} {\bibinfo  {journal} {arXiv
  preprint arXiv:2305.02907}\ } (\bibinfo {year} {2023})}\BibitemShut {NoStop}%
\bibitem [{\citenamefont {Caldwell}\ \emph {et~al.}(2018)\citenamefont
  {Caldwell}, \citenamefont {Didier}, \citenamefont {Ryan}, \citenamefont
  {Sete}, \citenamefont {Hudson}, \citenamefont {Karalekas}, \citenamefont
  {Manenti}, \citenamefont {da~Silva}, \citenamefont {Sinclair}, \citenamefont
  {Acala} \emph {et~al.}}]{caldwell2018parametrically}%
  \BibitemOpen
  \bibfield  {author} {\bibinfo {author} {\bibfnamefont {S.}~\bibnamefont
  {Caldwell}}, \bibinfo {author} {\bibfnamefont {N.}~\bibnamefont {Didier}},
  \bibinfo {author} {\bibfnamefont {C.}~\bibnamefont {Ryan}}, \bibinfo {author}
  {\bibfnamefont {E.}~\bibnamefont {Sete}}, \bibinfo {author} {\bibfnamefont
  {A.}~\bibnamefont {Hudson}}, \bibinfo {author} {\bibfnamefont
  {P.}~\bibnamefont {Karalekas}}, \bibinfo {author} {\bibfnamefont
  {R.}~\bibnamefont {Manenti}}, \bibinfo {author} {\bibfnamefont
  {M.}~\bibnamefont {da~Silva}}, \bibinfo {author} {\bibfnamefont
  {R.}~\bibnamefont {Sinclair}}, \bibinfo {author} {\bibfnamefont
  {E.}~\bibnamefont {Acala}}, \emph {et~al.},\ }\bibfield  {title} {\bibinfo
  {title} {Parametrically activated entangling gates using transmon qubits},\
  }\href@noop {} {\bibfield  {journal} {\bibinfo  {journal} {Physical Review
  Applied}\ }\textbf {\bibinfo {volume} {10}},\ \bibinfo {pages} {034050}
  (\bibinfo {year} {2018})}\BibitemShut {NoStop}%
\bibitem [{\citenamefont {Reagor}\ \emph {et~al.}(2018)\citenamefont {Reagor},
  \citenamefont {Osborn}, \citenamefont {Tezak}, \citenamefont {Staley},
  \citenamefont {Prawiroatmodjo}, \citenamefont {Scheer}, \citenamefont
  {Alidoust}, \citenamefont {Sete}, \citenamefont {Didier}, \citenamefont
  {da~Silva} \emph {et~al.}}]{reagor2018demonstration}%
  \BibitemOpen
  \bibfield  {author} {\bibinfo {author} {\bibfnamefont {M.}~\bibnamefont
  {Reagor}}, \bibinfo {author} {\bibfnamefont {C.~B.}\ \bibnamefont {Osborn}},
  \bibinfo {author} {\bibfnamefont {N.}~\bibnamefont {Tezak}}, \bibinfo
  {author} {\bibfnamefont {A.}~\bibnamefont {Staley}}, \bibinfo {author}
  {\bibfnamefont {G.}~\bibnamefont {Prawiroatmodjo}}, \bibinfo {author}
  {\bibfnamefont {M.}~\bibnamefont {Scheer}}, \bibinfo {author} {\bibfnamefont
  {N.}~\bibnamefont {Alidoust}}, \bibinfo {author} {\bibfnamefont {E.~A.}\
  \bibnamefont {Sete}}, \bibinfo {author} {\bibfnamefont {N.}~\bibnamefont
  {Didier}}, \bibinfo {author} {\bibfnamefont {M.~P.}\ \bibnamefont
  {da~Silva}}, \emph {et~al.},\ }\bibfield  {title} {\bibinfo {title}
  {Demonstration of universal parametric entangling gates on a multi-qubit
  lattice},\ }\href@noop {} {\bibfield  {journal} {\bibinfo  {journal} {Science
  advances}\ }\textbf {\bibinfo {volume} {4}},\ \bibinfo {pages} {eaao3603}
  (\bibinfo {year} {2018})}\BibitemShut {NoStop}%
\bibitem [{\citenamefont {Berry}(1984)}]{berry1984quantal}%
  \BibitemOpen
  \bibfield  {author} {\bibinfo {author} {\bibfnamefont {M.~V.}\ \bibnamefont
  {Berry}},\ }\bibfield  {title} {\bibinfo {title} {Quantal phase factors
  accompanying adiabatic changes},\ }\href@noop {} {\bibfield  {journal}
  {\bibinfo  {journal} {Proceedings of the Royal Society of London. A.
  Mathematical and Physical Sciences}\ }\textbf {\bibinfo {volume} {392}},\
  \bibinfo {pages} {45} (\bibinfo {year} {1984})}\BibitemShut {NoStop}%
\bibitem [{\citenamefont {Martinis}\ and\ \citenamefont
  {Geller}(2014)}]{martinis2014fast}%
  \BibitemOpen
  \bibfield  {author} {\bibinfo {author} {\bibfnamefont {J.~M.}\ \bibnamefont
  {Martinis}}\ and\ \bibinfo {author} {\bibfnamefont {M.~R.}\ \bibnamefont
  {Geller}},\ }\bibfield  {title} {\bibinfo {title} {Fast adiabatic qubit gates
  using only $\sigma$ z control},\ }\href@noop {} {\bibfield  {journal}
  {\bibinfo  {journal} {Physical Review A}\ }\textbf {\bibinfo {volume} {90}},\
  \bibinfo {pages} {022307} (\bibinfo {year} {2014})}\BibitemShut {NoStop}%
\bibitem [{\citenamefont {Chow}\ \emph {et~al.}(2010)\citenamefont {Chow},
  \citenamefont {DiCarlo}, \citenamefont {Gambetta}, \citenamefont {Motzoi},
  \citenamefont {Frunzio}, \citenamefont {Girvin},\ and\ \citenamefont
  {Schoelkopf}}]{chow2010optimized}%
  \BibitemOpen
  \bibfield  {author} {\bibinfo {author} {\bibfnamefont {J.~M.}\ \bibnamefont
  {Chow}}, \bibinfo {author} {\bibfnamefont {L.}~\bibnamefont {DiCarlo}},
  \bibinfo {author} {\bibfnamefont {J.~M.}\ \bibnamefont {Gambetta}}, \bibinfo
  {author} {\bibfnamefont {F.}~\bibnamefont {Motzoi}}, \bibinfo {author}
  {\bibfnamefont {L.}~\bibnamefont {Frunzio}}, \bibinfo {author} {\bibfnamefont
  {S.~M.}\ \bibnamefont {Girvin}},\ and\ \bibinfo {author} {\bibfnamefont
  {R.~J.}\ \bibnamefont {Schoelkopf}},\ }\bibfield  {title} {\bibinfo {title}
  {Optimized driving of superconducting artificial atoms for improved
  single-qubit gates},\ }\href@noop {} {\bibfield  {journal} {\bibinfo
  {journal} {Physical Review A}\ }\textbf {\bibinfo {volume} {82}},\ \bibinfo
  {pages} {040305} (\bibinfo {year} {2010})}\BibitemShut {NoStop}%
\bibitem [{\citenamefont {Lucero}\ \emph {et~al.}(2010)\citenamefont {Lucero},
  \citenamefont {Kelly}, \citenamefont {Bialczak}, \citenamefont {Lenander},
  \citenamefont {Mariantoni}, \citenamefont {Neeley}, \citenamefont
  {O’Connell}, \citenamefont {Sank}, \citenamefont {Wang}, \citenamefont
  {Weides} \emph {et~al.}}]{lucero2010reduced}%
  \BibitemOpen
  \bibfield  {author} {\bibinfo {author} {\bibfnamefont {E.}~\bibnamefont
  {Lucero}}, \bibinfo {author} {\bibfnamefont {J.}~\bibnamefont {Kelly}},
  \bibinfo {author} {\bibfnamefont {R.~C.}\ \bibnamefont {Bialczak}}, \bibinfo
  {author} {\bibfnamefont {M.}~\bibnamefont {Lenander}}, \bibinfo {author}
  {\bibfnamefont {M.}~\bibnamefont {Mariantoni}}, \bibinfo {author}
  {\bibfnamefont {M.}~\bibnamefont {Neeley}}, \bibinfo {author} {\bibfnamefont
  {A.}~\bibnamefont {O’Connell}}, \bibinfo {author} {\bibfnamefont
  {D.}~\bibnamefont {Sank}}, \bibinfo {author} {\bibfnamefont {H.}~\bibnamefont
  {Wang}}, \bibinfo {author} {\bibfnamefont {M.}~\bibnamefont {Weides}}, \emph
  {et~al.},\ }\bibfield  {title} {\bibinfo {title} {Reduced phase error through
  optimized control of a superconducting qubit},\ }\href@noop {} {\bibfield
  {journal} {\bibinfo  {journal} {Physical Review A}\ }\textbf {\bibinfo
  {volume} {82}},\ \bibinfo {pages} {042339} (\bibinfo {year}
  {2010})}\BibitemShut {NoStop}%
\bibitem [{\citenamefont {Motzoi}\ \emph {et~al.}(2009)\citenamefont {Motzoi},
  \citenamefont {Gambetta}, \citenamefont {Rebentrost},\ and\ \citenamefont
  {Wilhelm}}]{motzoi2009simple}%
  \BibitemOpen
  \bibfield  {author} {\bibinfo {author} {\bibfnamefont {F.}~\bibnamefont
  {Motzoi}}, \bibinfo {author} {\bibfnamefont {J.~M.}\ \bibnamefont
  {Gambetta}}, \bibinfo {author} {\bibfnamefont {P.}~\bibnamefont
  {Rebentrost}},\ and\ \bibinfo {author} {\bibfnamefont {F.~K.}\ \bibnamefont
  {Wilhelm}},\ }\bibfield  {title} {\bibinfo {title} {Simple pulses for
  elimination of leakage in weakly nonlinear qubits},\ }\href@noop {}
  {\bibfield  {journal} {\bibinfo  {journal} {Physical review letters}\
  }\textbf {\bibinfo {volume} {103}},\ \bibinfo {pages} {110501} (\bibinfo
  {year} {2009})}\BibitemShut {NoStop}%
\bibitem [{\citenamefont {Nguyen}\ \emph {et~al.}(2019)\citenamefont {Nguyen},
  \citenamefont {Lin}, \citenamefont {Somoroff}, \citenamefont {Mencia},
  \citenamefont {Grabon},\ and\ \citenamefont {Manucharyan}}]{nguyen2019high}%
  \BibitemOpen
  \bibfield  {author} {\bibinfo {author} {\bibfnamefont {L.~B.}\ \bibnamefont
  {Nguyen}}, \bibinfo {author} {\bibfnamefont {Y.-H.}\ \bibnamefont {Lin}},
  \bibinfo {author} {\bibfnamefont {A.}~\bibnamefont {Somoroff}}, \bibinfo
  {author} {\bibfnamefont {R.}~\bibnamefont {Mencia}}, \bibinfo {author}
  {\bibfnamefont {N.}~\bibnamefont {Grabon}},\ and\ \bibinfo {author}
  {\bibfnamefont {V.~E.}\ \bibnamefont {Manucharyan}},\ }\bibfield  {title}
  {\bibinfo {title} {High-coherence fluxonium qubit},\ }\href@noop {}
  {\bibfield  {journal} {\bibinfo  {journal} {Physical Review X}\ }\textbf
  {\bibinfo {volume} {9}},\ \bibinfo {pages} {041041} (\bibinfo {year}
  {2019})}\BibitemShut {NoStop}%
\bibitem [{\citenamefont {Ficheux}\ \emph {et~al.}(2021)\citenamefont
  {Ficheux}, \citenamefont {Nguyen}, \citenamefont {Somoroff}, \citenamefont
  {Xiong}, \citenamefont {Nesterov}, \citenamefont {Vavilov},\ and\
  \citenamefont {Manucharyan}}]{ficheux2021fast}%
  \BibitemOpen
  \bibfield  {author} {\bibinfo {author} {\bibfnamefont {Q.}~\bibnamefont
  {Ficheux}}, \bibinfo {author} {\bibfnamefont {L.~B.}\ \bibnamefont {Nguyen}},
  \bibinfo {author} {\bibfnamefont {A.}~\bibnamefont {Somoroff}}, \bibinfo
  {author} {\bibfnamefont {H.}~\bibnamefont {Xiong}}, \bibinfo {author}
  {\bibfnamefont {K.~N.}\ \bibnamefont {Nesterov}}, \bibinfo {author}
  {\bibfnamefont {M.~G.}\ \bibnamefont {Vavilov}},\ and\ \bibinfo {author}
  {\bibfnamefont {V.~E.}\ \bibnamefont {Manucharyan}},\ }\bibfield  {title}
  {\bibinfo {title} {Fast logic with slow qubits: microwave-activated
  controlled-z gate on low-frequency fluxoniums},\ }\href@noop {} {\bibfield
  {journal} {\bibinfo  {journal} {Physical Review X}\ }\textbf {\bibinfo
  {volume} {11}},\ \bibinfo {pages} {021026} (\bibinfo {year}
  {2021})}\BibitemShut {NoStop}%
\bibitem [{\citenamefont {Ding}\ \emph {et~al.}(2023)\citenamefont {Ding},
  \citenamefont {Hays}, \citenamefont {Sung}, \citenamefont {Kannan},
  \citenamefont {An}, \citenamefont {Di~Paolo}, \citenamefont {Karamlou},
  \citenamefont {Hazard}, \citenamefont {Azar}, \citenamefont {Kim} \emph
  {et~al.}}]{ding2023high}%
  \BibitemOpen
  \bibfield  {author} {\bibinfo {author} {\bibfnamefont {L.}~\bibnamefont
  {Ding}}, \bibinfo {author} {\bibfnamefont {M.}~\bibnamefont {Hays}}, \bibinfo
  {author} {\bibfnamefont {Y.}~\bibnamefont {Sung}}, \bibinfo {author}
  {\bibfnamefont {B.}~\bibnamefont {Kannan}}, \bibinfo {author} {\bibfnamefont
  {J.}~\bibnamefont {An}}, \bibinfo {author} {\bibfnamefont {A.}~\bibnamefont
  {Di~Paolo}}, \bibinfo {author} {\bibfnamefont {A.~H.}\ \bibnamefont
  {Karamlou}}, \bibinfo {author} {\bibfnamefont {T.~M.}\ \bibnamefont
  {Hazard}}, \bibinfo {author} {\bibfnamefont {K.}~\bibnamefont {Azar}},
  \bibinfo {author} {\bibfnamefont {D.~K.}\ \bibnamefont {Kim}}, \emph
  {et~al.},\ }\bibfield  {title} {\bibinfo {title} {High-fidelity,
  frequency-flexible two-qubit fluxonium gates with a transmon coupler},\
  }\href@noop {} {\bibfield  {journal} {\bibinfo  {journal} {arXiv preprint
  arXiv:2304.06087}\ } (\bibinfo {year} {2023})}\BibitemShut {NoStop}%
\bibitem [{\citenamefont {Barends}\ \emph {et~al.}(2019)\citenamefont
  {Barends}, \citenamefont {Quintana}, \citenamefont {Petukhov}, \citenamefont
  {Chen}, \citenamefont {Kafri}, \citenamefont {Kechedzhi}, \citenamefont
  {Collins}, \citenamefont {Naaman}, \citenamefont {Boixo}, \citenamefont
  {Arute} \emph {et~al.}}]{barends2019diabatic}%
  \BibitemOpen
  \bibfield  {author} {\bibinfo {author} {\bibfnamefont {R.}~\bibnamefont
  {Barends}}, \bibinfo {author} {\bibfnamefont {C.}~\bibnamefont {Quintana}},
  \bibinfo {author} {\bibfnamefont {A.}~\bibnamefont {Petukhov}}, \bibinfo
  {author} {\bibfnamefont {Y.}~\bibnamefont {Chen}}, \bibinfo {author}
  {\bibfnamefont {D.}~\bibnamefont {Kafri}}, \bibinfo {author} {\bibfnamefont
  {K.}~\bibnamefont {Kechedzhi}}, \bibinfo {author} {\bibfnamefont
  {R.}~\bibnamefont {Collins}}, \bibinfo {author} {\bibfnamefont
  {O.}~\bibnamefont {Naaman}}, \bibinfo {author} {\bibfnamefont
  {S.}~\bibnamefont {Boixo}}, \bibinfo {author} {\bibfnamefont
  {F.}~\bibnamefont {Arute}}, \emph {et~al.},\ }\bibfield  {title} {\bibinfo
  {title} {Diabatic gates for frequency-tunable superconducting qubits},\
  }\href@noop {} {\bibfield  {journal} {\bibinfo  {journal} {Physical review
  letters}\ }\textbf {\bibinfo {volume} {123}},\ \bibinfo {pages} {210501}
  (\bibinfo {year} {2019})}\BibitemShut {NoStop}%
\bibitem [{\citenamefont {Premaratne}\ \emph {et~al.}(2019)\citenamefont
  {Premaratne}, \citenamefont {Yeh}, \citenamefont {Wellstood},\ and\
  \citenamefont {Palmer}}]{premaratne2019implementation}%
  \BibitemOpen
  \bibfield  {author} {\bibinfo {author} {\bibfnamefont {S.~P.}\ \bibnamefont
  {Premaratne}}, \bibinfo {author} {\bibfnamefont {J.-H.}\ \bibnamefont {Yeh}},
  \bibinfo {author} {\bibfnamefont {F.}~\bibnamefont {Wellstood}},\ and\
  \bibinfo {author} {\bibfnamefont {B.}~\bibnamefont {Palmer}},\ }\bibfield
  {title} {\bibinfo {title} {Implementation of a generalized controlled-not
  gate between fixed-frequency transmons},\ }\href@noop {} {\bibfield
  {journal} {\bibinfo  {journal} {Physical Review A}\ }\textbf {\bibinfo
  {volume} {99}},\ \bibinfo {pages} {012317} (\bibinfo {year}
  {2019})}\BibitemShut {NoStop}%
\bibitem [{\citenamefont {O’Malley}\ \emph {et~al.}(2015)\citenamefont
  {O’Malley}, \citenamefont {Kelly}, \citenamefont {Barends}, \citenamefont
  {Campbell}, \citenamefont {Chen}, \citenamefont {Chen}, \citenamefont
  {Chiaro}, \citenamefont {Dunsworth}, \citenamefont {Fowler}, \citenamefont
  {Hoi} \emph {et~al.}}]{omalley2015}%
  \BibitemOpen
  \bibfield  {author} {\bibinfo {author} {\bibfnamefont {P.}~\bibnamefont
  {O’Malley}}, \bibinfo {author} {\bibfnamefont {J.}~\bibnamefont {Kelly}},
  \bibinfo {author} {\bibfnamefont {R.}~\bibnamefont {Barends}}, \bibinfo
  {author} {\bibfnamefont {B.}~\bibnamefont {Campbell}}, \bibinfo {author}
  {\bibfnamefont {Y.}~\bibnamefont {Chen}}, \bibinfo {author} {\bibfnamefont
  {Z.}~\bibnamefont {Chen}}, \bibinfo {author} {\bibfnamefont {B.}~\bibnamefont
  {Chiaro}}, \bibinfo {author} {\bibfnamefont {A.}~\bibnamefont {Dunsworth}},
  \bibinfo {author} {\bibfnamefont {A.}~\bibnamefont {Fowler}}, \bibinfo
  {author} {\bibfnamefont {I.-C.}\ \bibnamefont {Hoi}}, \emph {et~al.},\
  }\bibfield  {title} {\bibinfo {title} {Qubit metrology of ultralow phase
  noise using randomized benchmarking},\ }\href@noop {} {\bibfield  {journal}
  {\bibinfo  {journal} {Physical Review Applied}\ }\textbf {\bibinfo {volume}
  {3}},\ \bibinfo {pages} {044009} (\bibinfo {year} {2015})}\BibitemShut
  {NoStop}%
\end{thebibliography}%

\end{document}